\documentclass[a4paper,twocolumn, 11pt, accepted=2026-06-15]{quantumarticle}
\pdfoutput=1
\usepackage[numbers, sort&compress]{natbib}
\usepackage{amsmath} 
\usepackage{amssymb}
\usepackage{bm} 
\usepackage[retainorgcmds]{IEEEtrantools}
\usepackage{graphicx}
\usepackage{mathrsfs}
\usepackage{amsfonts}
\usepackage{tikz}
\usepackage{hyperref}
\usepackage{lipsum}
\usepackage{soul}
\usepackage{physics} 
\usepackage{float}
\usepackage{dsfont} 
\usepackage{mathrsfs}
\usepackage{BOONDOX-ds}
\usepackage{marvosym}
\usepackage{nicefrac}
\usepackage{bbold}
\usepackage[version=4]{mhchem}
\usepackage[english]{babel}

\newcommand{\bop}{{\textsf B}}
\newcommand{\E}[1]{\langle#1\rangle}

\title{Active Quantum Reservoir Engineering: Using a Qubit to Manipulate its Environment}
\author{Marcelo Janovitch} 
\email{m.janovitch@unibas.ch} 
\affiliation{Department of Physics and Swiss Nanoscience Institute, University of Basel, Klingelbergstrasse 82, 4056 Basel,
Switzerland }
\author{Matteo Brunelli}
\affiliation{JEIP, UAR 3573 CNRS, Coll\`ege de France, PSL Research University,
11 Place Marcelin Berthelot, 75321 Paris Cedex 05, France}
\affiliation{Department of Physics and Swiss Nanoscience Institute, University of Basel, Klingelbergstrasse 82, 4056 Basel,
Switzerland }
\author{Patrick P. Potts} 
\affiliation{Department of Physics and Swiss Nanoscience Institute, University of Basel, Klingelbergstrasse 82, 4056 Basel,
Switzerland }

\begin{document}
\maketitle
\begin{abstract}
 Quantum reservoir engineering leverages dissipative processes to achieve desired behaviour, with applications ranging from entanglement generation to quantum error correction.
 Therein, a structured environment acts as an entropy sink for the system, and no time-dependent control over the system is required. We develop a theoretical framework for \textit{active} reservoir engineering, where time-dependent control over a quantum system is used to manipulate its environment. In this case, the system may act as an entropy sink for the environment. Our framework captures the dynamical interplay between system and environment, and provides an intuitive picture of how finite-size effects and system-environment correlations allow for manipulating the environment by repeated initialisation of the quantum system. We illustrate our results with two examples: a superconducting qubit coupled to an environment of two-level systems and a semiconducting quantum dot coupled to nuclear spins. In both scenarios, we find qualitative agreement with previous experimental results, illustrating how active control can unlock new functionalities in open quantum systems.
\end{abstract}

\begin{figure*}[t]
\centering
\includegraphics[width=\linewidth]{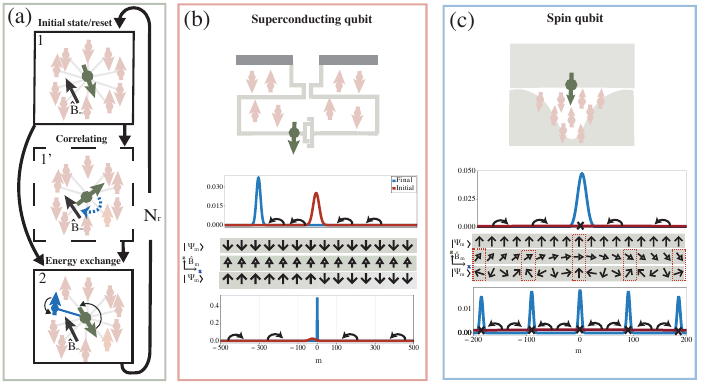}
\caption{Actively engineering the environment of a qubit. (a) Cyclic operation implementing active reservoir engineering. After initialising the qubit (step 1), it may be correlated with the environment (step 1') before exchanging energy (step 2). These steps are repeated multiple times to manipulate the magnetisation of the environment. (b) Active reservoir engineering using a superconducting qubit. Probability of the reservoir's magnetisation before (red) and after (blue) active reservoir engineering are shown for two different protocols, together with the Bloch vectors for the state after initialisation (arrows with open heads) and the direction of the magnetic field $\vec{B}_m$ (arrows with closed heads). Whenever the Bloch vector points in the same (opposite) direction as $\vec{B}_m$, the magnetisation increases (decreases), as illustrated by the curved arrows. The middle panel shows how repeated preparation of the state $\ket{\downarrow_z}$ results in cooling of the environment. The bottom illustrates how a correlated state can be used for cooling the environment if it has positive $m$ and heating it for negative $m$, resulting in a narrowing of the magnetisation. 
(c) Active reservoir engineering using a spin qubit. The middle illustrates how repeatedly preparing the state $\ket{\uparrow_z}$ results in narrowing of the magnetisation, because the field $\vec{B}_m$ rotates. The bottom shows how a correlated state can be used to create periodic narrowing, resulting in a distribution with multiple peaks. Dotted red boxes denote attractors of the magnetisation, where the state's Bloch vector is perpendicular to the field. Here $\Delta=0$.
For the superconducting qubit, we consider $N_r = 10^3$ iterations, and for the spin qubit $N_r=10^6$.}
\label{fig:overview} \end{figure*}

\section{Introduction}\label{s:intro}
Quantum systems inevitably couple to their surroundings. This is the virtue that allows them to be observed and the chagrin that ruins their unique quantum properties. To minimise unwanted noise and decoherence, they are typically kept as isolated from their environment as possible. 
In recent decades, dissipative processes have been recognised not only as a source of decoherence but also as a resource. This insight underpins quantum reservoir engineering~\cite{poyatos1996, Verstraete2009, Harrington2022}, where controlled dissipation enables desired functionality, such as non-reciprocity~\cite{Metelman2015}, topology~\cite{Bardyn2013} and autonomous quantum error correction~\cite{Leghtas2015, Lescanne2020, Gertler2021, Xu2023}, or stabilises resourceful states such as entangled~\cite{Kraus2008, Diehl2008, Barreiro2011, Lin2013, Leghtas2013, Zapletal2022} or squeezed states~\cite{Kronwald2013, wollman_2015, dassonneville_2021}. A key ingredient in reservoir engineering is a low-entropy environment, typically the vacuum state of the electromagnetic field. 

In many scenarios of interest, a low-entropy environment is not available because the relevant energy scales are comparable to temperature, rendering the standard approach to reservoir engineering inapplicable. This is typically the case in solid state qubits; the collective magnetisation of the material hosting the qubit can act as a bath, composed, e.g., of nuclear spins for spin qubits \cite{Urbaszek2013} or two-level defects for superconducting qubits~\cite{Klimov2018, Spiecker2024, Spiecker2024solomon}.  In this case, time-dependent control over the quantum system opens the door to a novel, \textit{active} approach to reservoir engineering, where the environment is manipulated through the system. Once the environment state is tailored, it can add functionality to the quantum system, acting as an engineered dissipative channel.

Active reservoir engineering protocols have already been experimentally implemented.
For instance, it has been shown that repeated resetting of a superconducting qubit to the ground state allows for cooling of the environment~\cite{Spiecker2024}. Furthermore, spin qubits in semiconductors have been extensively used to manipulate the magnetisation of their surrounding nuclei, also known as dynamic nuclear polarisation~\cite{Urbaszek2013}; this led to orders of magnitude improvement in coherence time \cite{Stepanenko2006, Maletinsky2006, Latta2009, Gangloff2019, Jackson2022, Nguyen2023, Shulman2014}. Compared to conventional (or passive) reservoir engineering, where the environment acts as an entropy sink and no time-dependent control over the system is required, the roles of system and environment are exchanged in active reservoir engineering: the system acts as an entropy sink for the environment.
Despite its potential, a unifying framework for active reservoir engineering remains largely undeveloped.

In this work, we fill this gap by deriving and analysing 
a theoretical framework for active reservoir engineering. This framework 
allows for an intuitive understanding of the joint system-bath dynamics and provides results that are in qualitative agreement with existing experiments.
It provides, to the best of our knowledge, the only tractable framework that captures finite-size effects and strong classical system-bath correlations, offering efficient simulations and analytic solutions where fully quantum methods are seldom tractable.
While we focus on specific platforms, our results apply to different central spin systems, including NV-centres~\cite{King2010, Khandelwal2023} and
we provide general derivations which can be used to treat systems beyond qubits.

\section{Overview}\label{s:over}

To illustrate our framework that is detailed in Sec.~\ref{s:mare}
, we focus on the two scenarios sketched in Fig~\ref{fig:overview}: a superconducting qubit coupled to a bath of two-level systems, motivated by the experiment in Ref.~\cite{Spiecker2024}, and a semiconducting quantum dot coupled to nuclear spins, motivated by Refs.~\cite{Jackson2022,Nguyen2023, Hogg2024}. In both cases, we consider a protocol where repeated initialisation of the system (henceforth called qubit) allows for manipulating the magnetisation, $m$, of the environment 
by altering its probability distribution $P_m$.
For a given magnetisation, the qubit will precess around an effective magnetic field $\vec{B}_m$, which depends on $m$. In addition, the qubit may exchange energy with the environment. This occurs through incoherent flip-flop processes, where the qubit and an environmental spin (or two-level system) flip together. Importantly, these processes are also dependent on the effective field $\vec{B}_m$. The Bloch vector of the qubit flips from being parallel to antiparallel with the magnetic field, or vice versa. If the qubit starts out (anti-)parallel to $\vec{B}_m$, then the flip-flop process increases (decreases) the magnetisation $m$ by one, c.f.~Fig.~\ref{fig:overview}\,(a). This provides an intuitive understanding of different strategies for active reservoir engineering: if the magnetisation should decrease, the qubit should be prepared with a Bloch vector that points in the direction opposite to $\vec{B}_m$. To increase the magnetisation, a state parallel to $\vec{B}_m$ should be prepared. 

\subsection{Superconducting qubit}
For the superconducting qubit, we consider an environment of two-level systems  (TLSs) that have approximately the same splitting as the qubit itself, such that energy exchanges between the system and environment are almost resonant. In this scenario, the effective field $\vec{B}_m$ is particularly simple --- it always points in the positive $z$-direction, see Fig.~\ref{fig:overview}\,(b). The magnetisation of the environment can thus be decreased by initialising the system in the spin-down state $\ket{\downarrow_z}$. Repeatedly initialising this state results in the environment being cooled, as it moves to lower energy states, see Fig.~\ref{fig:overview}. This protocol was employed in the experiment of Ref.~\cite{Spiecker2024}. Conversely, repeated initialisation in the spin-up state $\ket{\uparrow_z}$, results in population inversion (increased magnetisation) as also demonstrated experimentally in Ref.~\cite{Spiecker2024}.

Refined control of the state of the environment is possible by exploiting correlations between the system and the environment. As a simple example, consider the state $\ket{\downarrow_z}$ for $m>0$ and $\ket{\uparrow_z}$ for $m<0$. In this case, the magnetisation is decreased for positive $m$ and increased for negative $m$, resulting in a concentration of population in the $m=0$ state, as illustrated in Fig.~\ref{fig:overview}\,(b). While this correlated state may be difficult to prepare, it illustrates the potential of exploiting correlations to manipulate the state of the environment.

The superconducting qubit scenario, together with the protocols introduced here, is discussed in detail in Sec.~\ref{s:sc}.

\subsection{Quantum dot spin qubit}
In the quantum dot spin qubit scenario, there is a large mismatch between the energy splitting of the qubit and the nuclear spins that constitute the environment~\cite{Klauser2006, Coish2008decay, Jackson2022, Nguyen2023}. To enable energy transfer between the system and the environment, a Rabi drive with strength $\Omega$ is applied close to resonance with the nuclei frequencies (Hartmann-Hahn resonance). This has two important consequences. First, the energy exchange between the system and the environment can be switched off by removing the Rabi drive. Second, the direction of the field $\vec{B}_m = (\Delta-m|A_{\rm c}|)\hat{e}_z+\Omega\hat{e}_x$ depends on the magnetisation of the environment as illustrated in Fig.~\ref{fig:overview}\,c). Here $\Delta$ is the detuning of the drive and $A_{\rm c}$ is a coupling strength. For large positive (negative) magnetisation, the field points down (up) in the $z$-direction. In between it rotates, pointing in the $x$-direction at $m|A_{\rm c}|=\Delta$. This $m$-dependence of the field direction implies that if we repeatedly prepare the state $\ket{\uparrow_z}$, the magnetisation will decrease (increase) for large positive (negative) $m$, resulting in a concentration of population around $m=0$. This protocol has been implemented experimentally in Refs.~\cite{Nguyen2023, Hogg2024},  increasing the decoherence time of the spin qubit by one order of magnitude.

As for the superconducting qubit, further control of the bath is enabled by exploiting correlations. In the quantum dot spin qubit, a way to create such correlations is to switch off the Rabi drive and use the free evolution in the $m$-dependent field in a Ramsey protocol; the qubit is initialised in the $xy$-plane and precesses freely for a given time, with a final rotation to the $xz$-plane. In this manner, a state can be prepared with a Bloch vector whose angle
in the $xz$-plane is a function of $m$, see Fig.~\ref{fig:overview}\,(c). This state will result in periodic increase and decrease of the magnetisation as a function of $m$, creating peaks in its distribution by repeated preparation. A distribution with such qualitative features was observed experimentally in Ref.~\cite{Jackson2022}. Furthermore, as the locations of the peaks depend on the duration of the Ramsey protocol, all peaks except the one at $m=0$ can be suppressed by changing this duration from preparation to preparation. This protocol was implemented in Refs.~\cite{Jackson2022, Nguyen2023, Hogg2024}, increasing the coherence time by two orders of magnitude.

The quantum dot spin qubit scenario and the protocols discussed here are presented in more detail in Sec.~\ref{s:sq}.

\section{Master equation for active reservoir engineering}
\label{s:mare}
To describe active reservoir engineering protocols, we derive a master equation that keeps track not only of the system, but also the observable of interest in the environment, called master equation for active reservoir engineering (MARE). This is achieved through the technique of correlated projectors \cite{Esposito2003, Fischer2007, Breuer2006, Breuer2007, Sarma2012, Riera2021, Riera2022}. In App.~\ref{a:deriv} we provide  detailed derivation for an arbitrary quantum system. In contrast to Refs.~\cite{Riera2021, Riera2022} we treat part of the system-environment interaction non-perturbatively, leading to a more general master equation with qualitatively different features.

We focus on central spin models described by the Hamiltonian,
\begin{equation}
    \label{eq:hamiltonian}
    H = \vec{B}_0\cdot\vec{S}+\sum_{k=1}^N  \omega_k I_z^k+\sum_{k=1}^N\vec{S}\bm{A}\vec{I}_k,
\end{equation}
where $\vec{S} = (S_x, S_y, S_z)$ are spin-$\nicefrac{1}{2}$ operators describing the system qubit, $\vec{I}_k = (I^k_x, I^k_y, I^k_z)$ are spin-$s$ operators describing the environment, $\omega_k$ the frequencies of each bath constituent, and $\bm{A}$ is a $3\times 3$ coupling matrix describing  an anisotropic hyperfine contact interaction. For simplicity, we consider that each spin in the environment couples identically to the system. Inhomogeneous couplings are considered in App.~\ref{a:central-spin-mare}, where we specialise the MARE to a large class of central spin systems. We note that Eq.~\eqref{eq:hamiltonian} is beyond the integrable Gaudin magnets, due to the anisotropy in the coupling~\cite{Gaudin1976, Gareju2002, Schliemann2003, Tsyplyatyev2011, Sarma2012, He2022}
—the full quantum mechanical treatment of the bath thus typically relies on costly numeric simulations~\cite{Ruskuc2022, Zaporski2023, appel2024}.

The MARE derived from Eq.~\eqref{eq:hamiltonian} provides the joint dynamics for the qubit and the magnetisation of the environment, which is defined as the eigenvalue of the collective spin operator $J_z = \sum_k I^k_z$. The joint state of the qubit and the magnetisation is denoted $\rho_m(t)$.
The system density matrix can be obtained via $\rho_S(t) = \sum_m \rho_m(t)$ and the probability distribution of the magnetisation through $\tr\rho_m = P_m$.
The MARE reads
\begin{widetext}
\begin{align}\label{genMARE} 
\partial_t\rho_{m} = -i\qty[\vec{B}_m\cdot\vec{S},
\rho_{m}] +&\Gamma_{m}\qty[\dyad{\downarrow_m}{\uparrow_{m-1}}\frac{\rho_{m-1}}{V_{m-1}}
\dyad{\uparrow_{m-1}}{\downarrow_m} -\frac{1}{2}\qty{\dyad{\downarrow_m}{\downarrow_m}, \frac{\rho_{m}}{V_{m}}}]\\
+&\Gamma_{m+1}\qty[
\dyad{\uparrow_m}{\downarrow_{m+1}}\frac{\rho_{m+1}}{V_{m+1}}
\dyad{\downarrow_{m+1}}{\uparrow_{m}} -\frac{1}{2}\qty{\dyad{\uparrow_m}{\uparrow_m} ,
\frac{\rho_{m}}{V_{m}}}],\nonumber
\end{align}
\end{widetext}
where the state $\ket{\uparrow_m}$ ($\ket{\downarrow_m}$) corresponds to a Bloch vector that points in the same (opposite) direction as $\vec{B}_m$. The volume factors, $V_m$, give the number of states with a given magnetisation (i.e., the number of eigenstates of $J_z$ with eigenvalue $m$). 
Our MARE has the general form shown in Ref.~\cite{Breuer2007} to describe non-Markovian dynamics of $\rho_S$.

The first term in Eq.~\eqref{genMARE} corresponds to the unitary rotation of the qubit around the field $\vec{B}_m$. The second term corresponds to the jumps that increase the magnetisation by flipping the system spin down (along the axis defined by $\vec{B}_m$), and the third term describes the jumps that flip the system spin up, decreasing the magnetisation. Since a spin flip in the system is always accompanied by a change of magnetisation by one, the total magnetisation
\begin{equation}
    \label{totalM}
    M = m + \frac{1}{2}(\dyad{\uparrow_m}-\dyad{\downarrow_m}),
\end{equation}
is a conserved quantity. This implies that each time the system is prepared in an initial state, the magnetisation of the environment can change at most by one. When using a single qubit for active reservoir engineering, an iterative protocol is thus crucial to obtain a substantial manipulation of the environment, as depicted in Fig.~\ref{fig:overview}. Similar to Ref.~\cite{Riera2021}, the conserved quantity $M$ can be leveraged to solve the MARE analytically.
Akin to a Lindblad master equation in the secular approximation, the diagonal and the off-diagonal elements of $\partial_t\rho_m$  in the basis $\{\ket{\sigma_m}\},~\sigma = \uparrow,\downarrow$  decouple, and the populations evolve as,
\begin{align}
    \partial_t p(\downarrow, m)  &= \Gamma_m\qty[\frac{p(\uparrow,m-1)}{V_{m-1}} - \frac{p(\downarrow, m)}{V_m}]\label{pdwm},\\
    \partial_t p(\uparrow, m)  &= \Gamma_{m+1}\qty[\frac{p(\downarrow,m+1)}{V_{m+1}} - \frac{p(\uparrow, m)}{V_m}]\label{pupm},
\end{align}
where $p(\sigma, m) = \bra{\sigma_m}\rho_m\ket{\sigma_m}$.
The energy exchange between the system and the reservoir can thus be described by a classical rate equation for the joint probability, $p(\sigma,m)$.

To describe the scenarios illustrated in Fig.~\ref{fig:overview}, a state of the form
\begin{equation}
    \rho_m = P_m\dyad{\Psi_m}{\Psi_m},
\end{equation}
is prepared repeatedly and then time-evolved by Eq.~\eqref{genMARE}. For the uncorrelated states illustrated in Fig.~\ref{fig:overview}, $\ket{\Psi_m}=\ket{\downarrow_z}$ and $\ket{\Psi_m}=\ket{\uparrow_z}$ have been used for the superconducting qubit and the quantum dot spin qubit respectively. The correlated states in Fig.~\ref{fig:overview} are given in Eqs.~\eqref{thetastate} and \eqref{ramseystate} below, where the corresponding protocols are discussed in depth. To simulate multiple cycles, the distribution $P_m$ obtained at the end of a cycle is used as the initial distribution for the next cycle. 

\subsection{Thermodynamics of active reservoir engineering}
In active quantum reservoir engineering, entropy is typically removed from the environment through the qubit.
This process is constrained by the laws of thermodynamics.
At step $1$ in Fig.~\ref{fig:overview}\,(a) the qubit is initialised, typically in a pure state.
At step $2$, the qubit evolves together with the bath through the MARE. During this process, the entropy of the system-environment compound cannot decrease. However, an increase in the entropy of the qubit allows for the entropy of the environment to decrease. At step 1 of the next cycle, the qubit is reset reducing its entropy. Only at this stage can the entropy of the system-environment compound \textit{decrease}, as in a Szilard engine~\cite{Szilard1929, MaxwellDemons, Spiecker2024}.

For a quantitative analysis, we model the system-bath entropy using the \textit{observational entropy}~\cite{Safranek2019, Strasberg2021, Riera2021},
\begin{align}\label{Sobs}
    \mathcal{S}_\text{obs} = -\sum_{\sigma, m} p(\sigma, m)\ln p(\sigma, m) + \sum_{m} P_m \ln V_m.
\end{align}
The first term in Eq.~\eqref{Sobs} corresponds to the Shannon entropy of the joint probability $p(\sigma, m)$. It contains contributions of the marginal entropies, such as the bath Shannon entropy, as well as the mutual information~\cite{Riera2021}. The Shannon entropy of the bath,
\begin{align}\label{SB}
    \mathcal{S}_B = -\sum_m P_m \ln P_m
\end{align}
is central to engineering $P_m$---it quantifies the uncertainty about the magnetisation, $m$. The Shannon entropy of the remaining marginal can be related to the von Neumann entropy of the system (see App.~\ref{a:decomposition}),
\begin{align}\label{Svn}
    \mathcal{S}(\rho_S) = -\tr \rho_S\ln\rho_S.
\end{align}
The second term in Eq.~\eqref{Sobs} is the average of the Boltzmann entropy, $S^\text{Boltz}_m = \ln V_m$~\cite{Safranek2019}. 

The MARE ensures the second law of thermodynamics, see App.~\ref{a:second-law} for a derivation
\begin{align}\label{eq:second-law}
    \partial_t \mathcal{S}_\text{obs} \geq 0.
\end{align}
As the qubit entropy rises in step $2$, the second law limits how much the Shannon entropy of the bath can decrease.
In summary, the thermodynamic principles of quantum active reservoir engineering are the following. 
\begin{enumerate}
\item[(i)] Observational entropy is redistributed by system-environment interactions, consuming the qubit’s purity to lower the Shannon entropy of the bath. 
\item[(ii)] Repeated resetting of the qubit reduces the observational entropy bit-by-bit.
\end{enumerate}
\subsection{Assumptions}\label{ss:assumptions}
The MARE relies on a number of assumptions that we summarise here, see Apps.~\ref{a:deriv} and~\ref{a:central-spin-mare} for more details. In Section~\ref{ss:sq_validity} we show how the validity of these assumptions can be assessed for a concrete model and fixed parameters.

\textbf{(i) Identification of the bath observable of interest.} Equation~\eqref{eq:hamiltonian} has the general structure,
\begin{equation}
    \label{eq:hamiltonian_abstract}
    H = H_S + H_B + H_{SB},
\end{equation}
where, in particular, se have $H_S = \vec{B}_0\cdot\vec{S}$, $H_B=\sum_{k=1}^N  \omega_k I_z^k$, and $H_{SB} = \sum_{k=1}^N\vec{S}\bm{A}\vec{I}_k$.
Instead of treating the full quantum state of the reservoir, the MARE aims to describe the dynamics of a given bath observable, $M_B$, which satisfies $[H_B, M_B] = 0$. The MARE captures the evolution of the probability distribution of finding a given eigenstate of $M_B$, $m$, at a time $t$, namely $P_m(t)$.
Here, the observable of interest is the collective magnetisation $M_B = J_z$, satisfying $[M_B, H_B]=0$. Physically, for both superconducting- and spin-qubit models discussed, the magnetisation plays a pivot role in dissipation and decoherence originating from the solid-state substrate where those qubits are built. The formalism can be adapted to other choices, such as $M_B = \sum_k c_k I_z^k$, where $c_k$ are arbitrary.

\textbf{(ii) Perturbative and non-perturbative contributions.} The MARE treats perturbatively the parts of the interaction which do not commute with $M_B = J_z$. In Eq.~\eqref{eq:hamiltonian}, these are the terms proportional to $J_y$ and $J_x$. In contrast to previous approaches \cite{Riera2021, Riera2022}, however, we treat the part of the interaction proportional to $J_z$ \textit{non-perturbatively}. This distinctive feature of the MARE is particularly relevant to capture the dynamics over many cycles, when the magnetisation is substantially modified. Note that a different choice of $M_B$ implies a different split between what can be can treated perturbatively and what cannot.

\textbf{(iii) Markov approximation.} We require the environmental decoherence time to be short compared to the energy exchange between system and bath (i.e. compared to the rates, $\Gamma_m$). In our model, environmental decoherence arises due to the spread in the frequencies $\omega_k$ and their broadening due to extrinsic decoherence mechanisms, which are captured within a bath spectral density description. The validity of this approximation is thus directly related to the characteristic width of the bath spectral density. Hence, the MARE cannot capture any effect that requires quantum coherence in the environment, such as nuclear dark states \cite{appel2024, Cai2025} or the use of the environment as a quantum register \cite{Denning2019, appel2024}. 
While the latter are experiments in which bath coherence is relevant, decoherence in the reservoir is still the reality of many state-of-the-art experiments, and our approach provides a robust benchmark of its relevance.

\textbf{(iv) Secular approximation.} The MARE assumes that the splitting between the levels of $\vec{B}_m \cdot \vec{S}$ is much larger than the rates $\Gamma_m$ for all values of $m$.

\section{Superconducting qubit}\label{s:sc}
\subsection{System and model}
There is mounting  evidence that
superconducting qubits are surrounded by many two-level defects (TLSs) with
different frequencies~\cite{Wu2012, Shalibo2010, Quintana2017, Klimov2018, Yan2016, Spiecker2024}. This can be captured by the Hamiltonian,
\begin{align}\label{sc-total-ham} H = \omega_S S_z + \sum_{k=1}^N\omega_k S^k_z
+ V, \end{align} where the qubit (system) has frequency $\omega_S$, the sum runs
over the ensemble of TLSs and $S^{(k)}_z$ are spin-$\nicefrac{1}{2}$ operators. 

There are several conjectures about the origin of the TLS bath in the
superconducting circuit~\cite{Klimov2018, Spiecker2024}. Here, we consider that the bath is coupled to the qubit through a
Fermi-contact hyperfine interaction~\cite{Wu2012,Quintana2017},
\begin{align}\label{sc-interaction2} V
=  A\sum_{k=1}^N S_z S_z^k  + A\sum_{k=1}^N \qty(S_x S_x^k + S_y S_y^k) . \end{align}
We focus on cases in which
$V$ represents \textit{weak} coupling, with hyperfine constant $A$.  We also focus on quasi-resonant TLS, such that all frequencies $\omega_k$ are close to $\omega_S$. Since off-resonant TLSs cannot exchange energy with the superconducting qubit, but only contribute to its dephasing, our results are also relevant in the presence of off-resonant TLSs, see App.~\ref{a:SW} for details. Furthermore, we disregard the spin-flip processes on the TLSs which would compete to rethermalise the TLSs bath~\cite{Spiecker2024, Spiecker2024solomon}.

Based on the microscopic model,
we find the MARE [see App.~\ref{a:sc} for a derivation],
\begin{align}\label{sc-MARE}
&\partial_t\rho_{m} = -i\qty[(\omega_S + A m)S_z,
\rho_{m}] \\ &+ \kappa V_m\qty(\frac{1}{2}+\frac{m}{N})
\qty[\sigma \frac{\rho_{m-1}}{V_{m-1}} \sigma^\dagger
-\frac{1}{2}\qty{\sigma \sigma^\dagger,
\frac{\rho_{m}}{V_{m}}}]\nonumber\\ &+ \kappa
V_m\qty(\frac{1}{2}-\frac{m}{N})\qty[\sigma^\dagger
\frac{\rho_{m+1}}{V_{m+1}} \sigma -\frac{1}{2}\qty{\sigma^\dagger
\sigma, \frac{\rho_{m}}{V_{m}}}],\nonumber \end{align} 
where $\sigma = \dyad{\downarrow_z}{\uparrow_z}$, $\kappa$ denotes the system-bath coupling, and the volume factors are given by (assuming even $N$)~\cite{Salinas, Dicke1954},
\begin{equation}
    \label{vm}
    V_m =\frac{N!}{(\nicefrac{N}{2}+m)!(\nicefrac{N}{2}-m)!}.
\end{equation}
The equation above has the same form as Eq.~\eqref{genMARE}, with 
\begin{align}
\vec{B}_m = (\omega_S  + Am)\hat{e}_z.
\end{align}
Equation~\eqref{sc-MARE} also has the same form as the one derived in
\cite{Riera2021, Riera2022}. We note that for distributions $P_m$ with support that is far away from the ground state or the highest excited state ($m=\mp N/2$), the factors $m/N$ in the rates do not play a role. However, they ensure that the distribution stays within $m\in [-N/2,N/2]$  since $\Gamma_{-N/2}=\Gamma_{N/2+1}=0$.

Previous theoretical approaches using Solomon equations~\cite{Spiecker2024solomon, Somon1955} neglect coherence in the qubit and all system-bath correlations. Neglecting coherence is reasonable in the absence of a coherent drive. However, as we discuss below, correlations are a valuable resource in active reservoir engineering. 

\subsection{Cooling the TLS bath}\label{ss:cooling}
\begin{figure*}[t] \centering
\begin{tikzpicture}
\node (a)  at (0,0)
{\includegraphics[width=\linewidth]{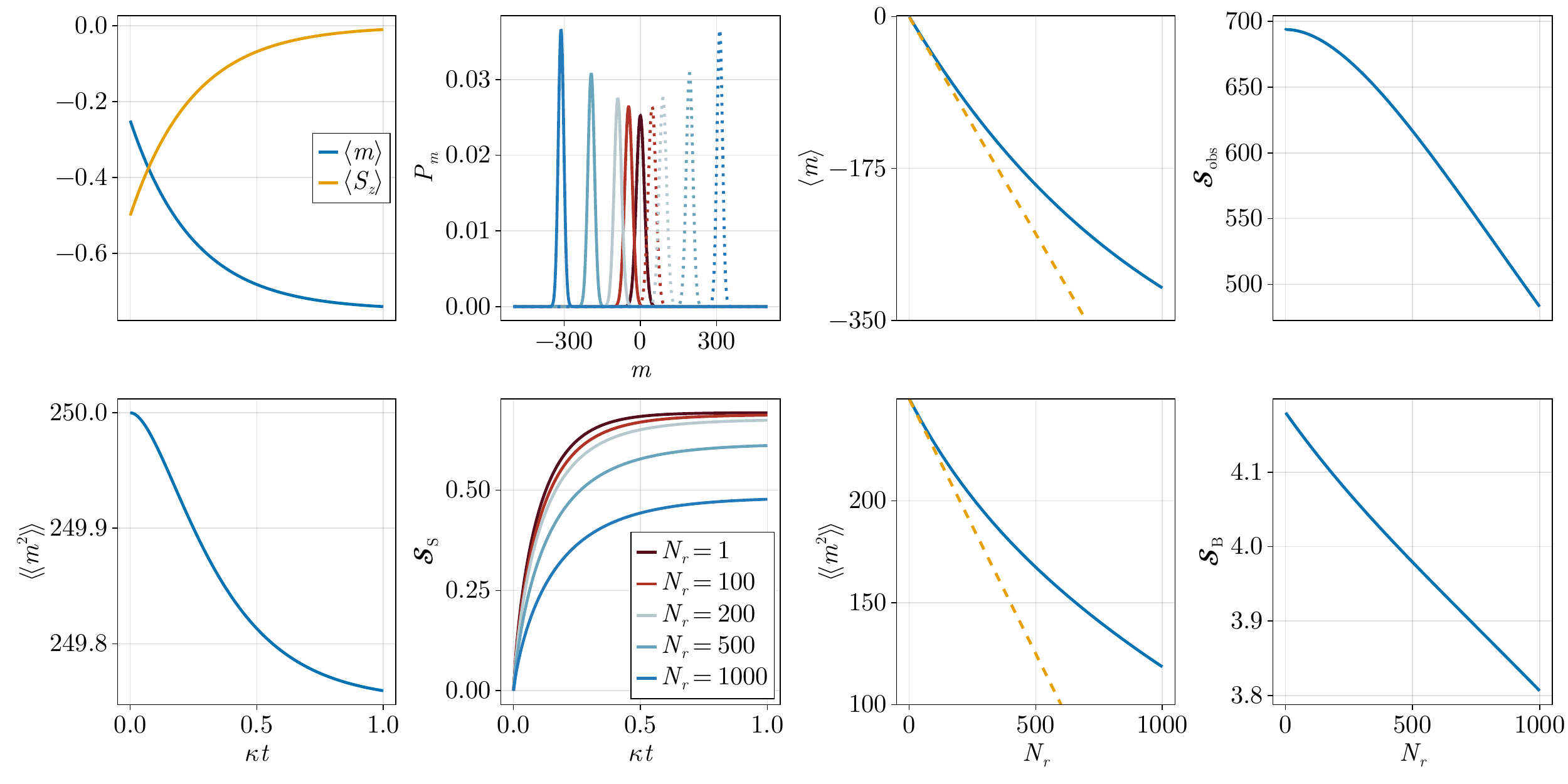}};
\node (a) at (-7.25,4.75) {\large(a)};
\node (b) at (-7.25,0.25) {\large(b)};
\node (c) at (-3, 4.75) {\large(c)};
\node (d) at (-3, 0.25) {\large(d)};
\node (e) at (1.7, 4.75) {\large(e)};
\node (f) at (1.7, 0.25) {\large(f)};
\node (g) at (6, 4.75) {\large(g)};
\node (h) at (6, 0.25) {\large(h)};
\end{tikzpicture}
\caption{
TLSs bath cooling. We consider a bath with
$N=1000$ TLSs and, $A/\omega_S = 0.1/N$, $\beta\omega_S = 0.001, \kappa/\omega_S = 10^{-5}$.  
(a) Change in bath (blue) and system (yellow) magnetisation in the first iteration of the cooling protocol.
(b) Decrease in the bath variance for the first iteration. (c) $P_m$ at the end of $N_r$ repetitions initialising the $\ket{\downarrow_z}$ state in the system. Dotted lines indicate population inversion protocol, in which we initialise $\ket{\uparrow_z}$ at each cycle. (d) Von Neumann entropy of the system during different iterations of the cooling protocol (c). We observe that the maximum entropy achieved decreases, witnessing the cooling of the reservoir through the qubit. (e) Average magnetisation by the end of each repetition. The yellow dashed curve represents the linear term in $N_r$, from Eq.~\eqref{sc-avg-cycles}. (f) Decrease in the variance, the yellow dashed is the linear term contribution in $N_r$ from Eq.~\eqref{sc-var-cycles}. (g) As the bath cools, $P_m$ peaks at less degenerate values of $m$, decreasing the observational entropy. The change in observational entropy contains relevant contributions from the change in the average Boltzmann entropy. (h) Decrease in the Shannon entropy of the bath due to narrowing of the distribution.
}
\label{fig:sc-panel} 
\end{figure*}
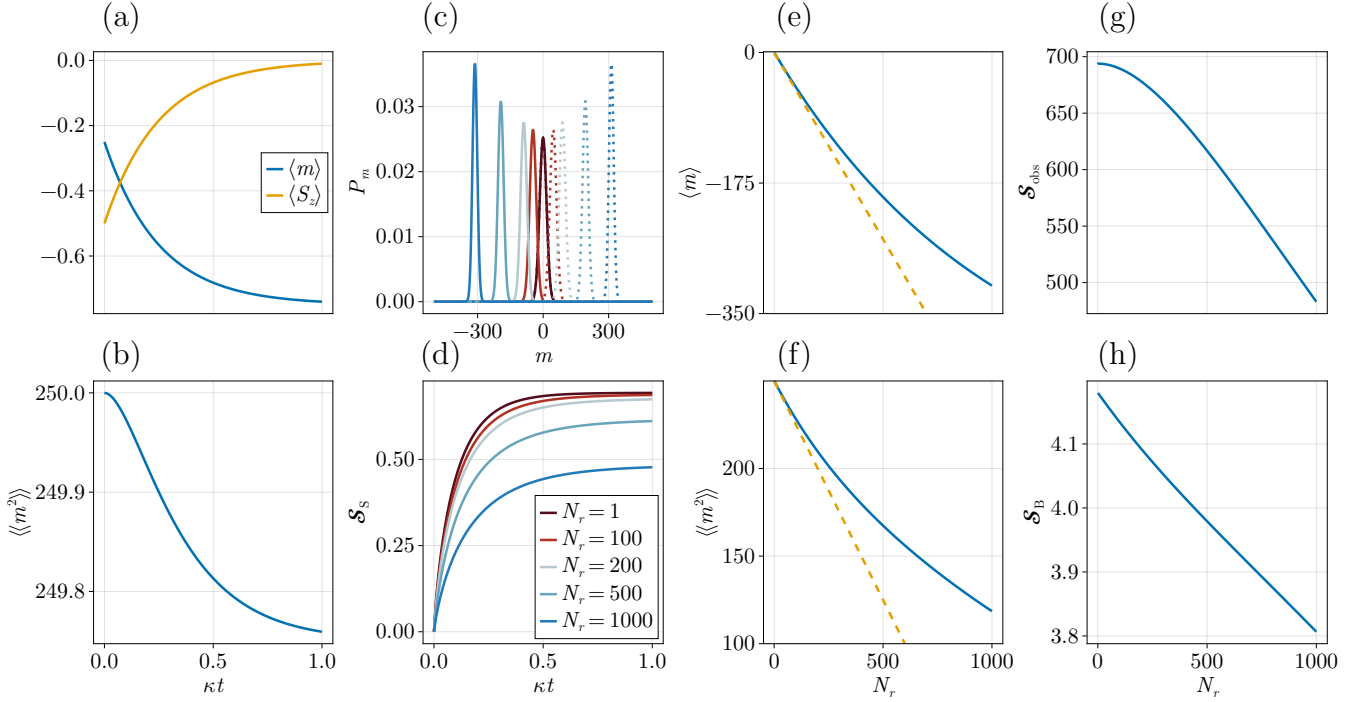
Assuming the bath is initially in a thermal state, we aim to bring it closer to the ground state. Since we focus on quasi-resonant TLSs, they all have an energy that is close to $\omega_S$ and the initial probability of magnetisation in a thermal state is well approximated by
 $P_m =e^{-\beta \omega_S m} V_m/Z$. 
If the number of quasi-resonant spins is big enough, $P_m$ is well approximated by a Gaussian with average $\mu_\beta$ and standard deviation $\varsigma_\beta$. The aim of the cooling protocol is to move the distribution $P_m$ towards smaller (or more negative) values of $m$. Ideal cooling would result in all TLSs in the ground state, which corresponds to $P_m = \delta_{m,-N/2}$.
We start by  investigating the equations of motion for the first and second moment
of $m$.

From the MARE~\eqref{sc-MARE} we can obtain the equation of motion for the
average, 
\begin{align} \partial_t\expval{m} = -\frac{\kappa}{N} \expval{m} + \kappa
\expval{S_z}, \label{sc-avg-eq}
\end{align}
 which is supplemented by $\partial_t \expval{M} =0$, where the conserved quantity is $M=m + S_z$. These equations can be solved. In particular, in the long-time limit we find 
\begin{align}
\expval{m}_\infty=\expval{M}\frac{N}{N+1},\hspace{.2cm}
\langle S_z\rangle_\infty = \frac{\langle M\rangle}{N+1}.
\label{eq:sc-mean-ss}
\end{align}
The long-time
behaviour is  determined by the conserved quantity and the number of TLSs, $N$.  

For the second moment, we find, 
\begin{align}
\partial_t\langle m^2\rangle = -\frac{2\kappa}{N} \langle m^2\rangle + 2\kappa \frac{N-1}{N} \expval{S_z m} + \frac{\kappa}{2}\label{sc-var-eq}, 
\end{align}
where we observe that it
couples to the correlator with the $z$-magnetisation of the system. We obtain for the variance in the long-time limit the
following expression, 
\begin{align}\label{sc-var-ss}
&\langle\!\langle m^2\rangle\!\rangle_\infty = 
\frac{N-1}{N+1}\langle \! \langle M^2 \rangle \!\rangle_0 -\frac{\langle M \rangle^2_0}{(N+1)^2} + \frac{1}{4},
\end{align} 
where we used $\expval{M^2}=\expval{m^2}+\nicefrac{1}{4}
+2\expval{S_z m}$, we introduced the notation $\langle\!\langle AB\rangle\!\rangle = \langle AB\rangle-\langle A\rangle\langle B\rangle$,
and the subscript $0$ denotes initial conditions.

To illustrate the
result, we consider in this Section the qubit to be initially in the ground state $\expval{S_z}_0 = -\nicefrac{1}{2}$, uncorrelated with the
bath which starts out in a thermal state. This results in
\begin{align}
\expval{m}_\infty
&=\langle m\rangle_\beta - \frac{1}{2} + \mathcal{O}(\nicefrac{1}{N}),\\\langle S_z\rangle_\infty &= \mathcal{O}(\nicefrac{1}{N}),
\end{align}
and
\begin{equation}
\langle\!\langle m^2\rangle\!\rangle_\infty =
\langle\!\langle m^2\rangle\!\rangle_\beta-\frac{1}{4} + \mathcal{O}(\nicefrac{1}{N}), 
\end{equation} 
{ where we used that for a thermal state the first and second cumulants are extensive prior to taking the large-$N$ limit. i.e., $\mu_\beta,~\langle\!\langle m^2\rangle\!\rangle_\beta~\propto N$ }
From this we can conclude that, up to the finite-size contributions, we flip a single
spin of the bath with probability one half; the variance decreases by $\nicefrac{1}{4}$ and the average drifts by a factor of $\nicefrac{1}{2}$. 
{ Note that this does not hold for general (non-thermal) initial states; a narrow initial bath magnetisation state with $\langle\!\langle m^2\rangle\!\rangle_0 = \mathcal{O}(1)$ and $\langle m\rangle _0 = 0$ under the same dynamics will \textit{increase} its variance.}

We now consider the following cyclic protocol, as sketched in Fig.~\ref{fig:overview}\,(a). The qubit is initialised in the ground state $|\!\downarrow_z\rangle$, i.e.
$\expval{S_z}_0=-\nicefrac{1}{2}$. 
Then, the system and bath interact for long enough that the steady-state is approached, i.e. $t\gg1/\kappa$. Finally, the qubit is reset, destroying any
system-bath correlations. This protocol is repeated $N_r$ times, as depicted in Fig.~\ref{fig:overview} (a). According to Eq.~\eqref{eq:sc-mean-ss} and \eqref{sc-var-ss}, each cycle decreases the average magnetisation by $1/2$ and its variance by $1/4$, until the finite size corrections that go as $N_r/N$ become important;
\begin{align}
\expval{m}^{(N_r)} &= \langle m \rangle_\beta - \frac{N_r}{2}
+\mathcal{O}(\nicefrac{N_r}{N}),\label{sc-avg-cycles}\\
\langle\!\langle m^2\rangle\!\rangle^{(N_r)} &= \langle\!\langle m^2 \rangle\!\rangle_\beta
-\frac{N_r}{4}+\mathcal{O}(\nicefrac{N_r}{N})\label{sc-var-cycles}. \end{align}
A sizeable change in the bath thus requires a number of cycles comparable to the bath size. This is illustrated in Fig.~\ref{fig:sc-panel}\,(e, f). A similar protocol initialising the state $|\!\uparrow_z\rangle$ results in population inversion instead of cooling. Both these scenarios have been recently observed experimentally~\cite{Spiecker2024}.

Figure~\ref{fig:sc-panel}~(d) shows the von Neumann entropy, Eq.~\eqref{Svn}
and Figs.~\ref{fig:sc-panel}~(g, h)  show  the Shannon, Eq.~\eqref{SB} and observational,~Eq.~\eqref{Sobs}, entropies 
by the end of each cycle in a cooling protocol. 
As expected, the Shannon entropy of the bath decreases. The observational entropy also decreases, but its decrease is steeper relative to the Shannon entropy; since $P_m$ drifts towards a less degenerate magnetisation, the contribution from the average Boltzmann entropy, $\sum_m P_m \ln V_m$, also decreases significantly.
The qubit thermalises with the bath at each cycle and can be regarded as a thermometer; for the first cycles, its entropy reaches $\approx \log 2$---the maximum for a qubit---, and as the bath cools the qubit reaches smaller entropies, thermalising to a state with lower temperature.

\subsection{Using correlations for engineering $P_m$} \label{ss:sc_correlated}

In this section, we investigate how a correlated state allows to narrow the bath distribution. We first consider an ideal correlated state, before turning to a class of states inspired by recent experiments on spin qubits. In this section, we do not consider the problem of creating the correlations. An example of how such correlations can be created in a different system is discussed below in Sec.~\ref{s:sq_corr}. {We concentrate  our analysis on Gaussian states $P_m \approx \mathcal{N}(\mu, \varsigma)$ with average $\mu=0$ and variance $\langle\!\langle m^2\rangle\!\rangle_0 =\varsigma^2$, such as the infinite temperature thermal state, with $\varsigma_{\beta = 0} = \sqrt{\nicefrac{N}{4}}$}.

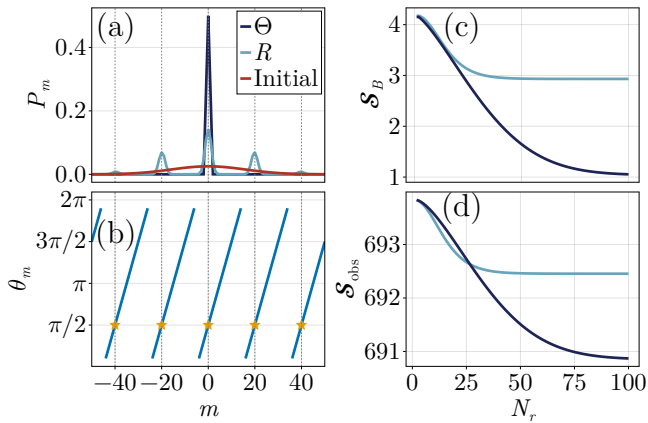
\begin{figure}[t]
    \centering
    \begin{tikzpicture}
    \node (pic) at (0,0) {\includegraphics[width=\linewidth]{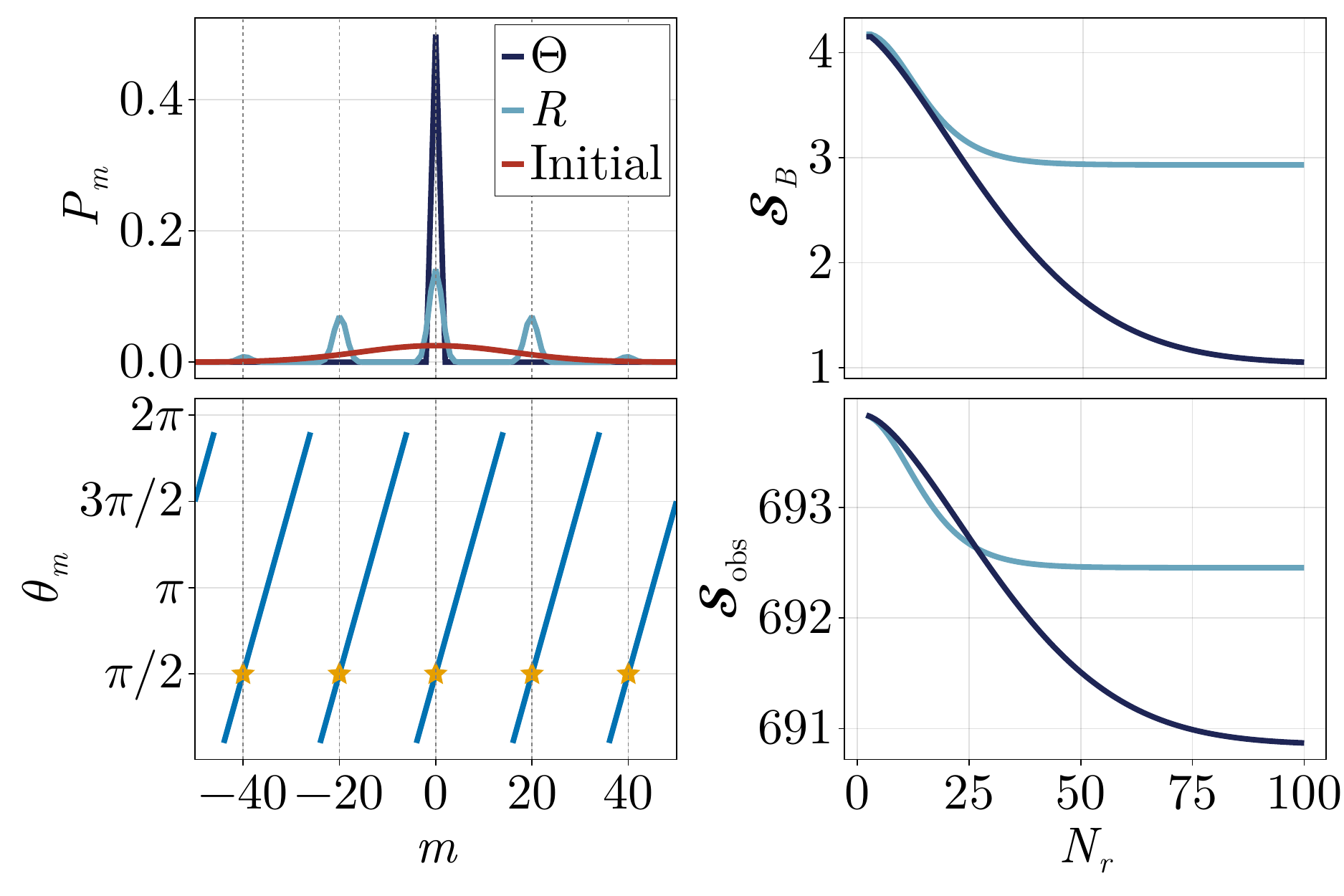}} ;
    \node (a) at (-2.5, 2.3) {\large(a)};
    \node (b) at (-2.6, -0.25) {\large(b)};
    \node (c) at (1.85, 2.25) {\large(c)};
    \node (d) at (1.85, 0.0) {\large(d)};
    \end{tikzpicture}
    \caption{TLSs reservoir engineering using correlated states. We consider $N=1000,~A/\omega_S = 0.1/N, ~\beta=0, \kappa/\omega_S = 10^{-5},$ and $\alpha = 2\pi/20,~ \varphi = \pi/2$ for the Ramsey state. (a) $P_m$ at after $N_r =100$ repetitions using each correlated state. We observe optimal narrowing for $\rho^\Theta$ and formation of satellite peaks in integer multiples of $20$ for $\rho^R$. (b) Angle between the initial Ramsey state's field and $\vec{B}_m$. The stars denote the points at which $\theta_m = \pi/2$, where peaks in $P_m$ are expected. (c) Shannon entropy of the bath by the end of each repetition. For the $\rho^\Theta$ we observe a huge decrease, and for $\rho^R$ the presence of peaks limits the decrease of entropy. (d) The change of observational entropy is dominated by the Shannon entropy since the distribution is always centred at $m=0$.}
    \label{fig:sc-purification}
\end{figure}

The jump terms in the MARE~\eqref{sc-MARE} indicate the ideal correlated state to achieve narrowing of $P_m$ around $m=0$:  for $m<0$ the system state is aligned with $\vec{B}_m$ and for $m>0$ it is anti-aligned.
Formally, this is described by the state (for $m\neq0$), $\rho_m^\Theta = P_m |\Psi_m^\Theta\rangle\langle\Psi_m^\Theta| $, with,
\begin{align}
\label{thetastate}
	|\Psi_m^\Theta\rangle = \Theta_{m} \ket{\downarrow_z} + (1-\Theta_{m}) \ket{\uparrow_z},
\end{align}
where $\Theta_m$ is the Heaviside step function which is $+1$ ($0$) for $m> 0$ ($m<0$).
For $m=0$, we set $\rho_0^\Theta = P_0 \left(\dyad{\uparrow_z}+\dyad{\downarrow_z}\right)/2.$
With this initial state we find 
{
\begin{align}
&\expval{S_z}_0=0,~\langle\!\langle S_z^2 \rangle \! \rangle_0=\frac{1}{4}, \label{eq:sc_correlated_Sz}\\
&\expval{m S_z}_0=-\frac{\langle|m|\rangle_0}{2}. \label{eq:sc_correlated_cov}
\end{align} }
For the considered $P_m$, Eq.~\eqref{eq:sc_correlated_cov} reduces to
\begin{align}\label{covideal}
\langle\!\langle S_z m\rangle\!\rangle_0=-\frac{\varsigma}{\sqrt{2\pi}}.
\end{align}
Equations~(\ref{eq:sc-mean-ss},
\ref{sc-var-ss}) then yield,
\begin{align}
	\expval{m}_\infty &= 0,\\
    \langle\!\langle m^2\rangle\!\rangle_\infty &=  
    {\varsigma^2\qty(1- \frac{2}{N}) 
    -\varsigma\sqrt{\frac{2}{\pi}} + \frac{1}{2} + \mathcal{O}\qty(\nicefrac{1}{N})},\label{eq:sc_var_reduction_Theta}
\end{align}
{where we assumed that $\varsigma^2 = \mathcal{O}(N)$}.
Therefore, the anti-correlated state narrows the
distribution proportionally to its standard deviation without polarising the bath. For example, if $N=10^4$
and $\varsigma=50$, then a single cycle of interaction with the correlated
state is comparable to the variance reduction of flipping $\approx80$ TLSs in the absence of correlations. As illustrated in Fig.~\ref{fig:sc-purification}, correlations allow for drastically narrowing the bath distribution by repeatedly initialising the state in Eq.~\eqref{thetastate}. Both the entropy of the bath as well as the variance of $P_m$ are drastically reduced already within the first 10 cycles, i.e. for $N_r\ll N$. We note that even with the ideal state given in Eq.~\eqref{thetastate}, it is not possible to perfectly narrow to $P_m = \delta_{m,0}$. The reason is that if the bath already is in the state $m=0$, there are still processes that take the bath to $m=\pm1$.
Despite their potential for narrowing, correlations cannot enhance the drift in the magnetisation of the bath (i.e., move its mean) as it still holds that at most one TLS can be flipped in each cycle.

We now  consider a different class of correlated states, the Ramsey-correlated state
$\rho_m^R = P_m |\Psi_m^R\rangle\langle\Psi_m^R| $, where
\begin{align} 
\label{ramsey-state-sc}
    |\Psi_m^R\rangle = \cos\frac{\theta_m}{2}
\ket{\uparrow_z} + \sin\frac{\theta_m}{2}
\ket{\downarrow_z},
\end{align} 
and we highlight that the above state is the same class of states used to produce the satellite peaks in Fig.~\ref{fig:overview}~(c).
We assume a linear phase $\theta_m =\alpha m+\varphi$, which can be optimised to prepare an
anti-correlated state. 

{ Equations~(\ref{eq:sc_correlated_Sz},~\ref{eq:sc_correlated_cov}) still hold for $\rho_m^R$, and the covariance for $P_m = \mathcal{N}(\mu=0,
\varsigma)$ can also be computed analytically. We find \textit{maximal} (negative) covariance of 
\begin{align}
\langle\!\langle S_z m\rangle\!\rangle_0= -\frac{\varsigma}{2\sqrt{e}},\label{sc-ramsey-correlations} 
\end{align}
by setting $\alpha=1/\varsigma$ and $\varphi=-\pi/2$.  }
The state in Eq.~\eqref{ramsey-state-sc} can be used to narrow the bath distribution, but it performs worse than the ideal case as we obtain,
\begin{align}
    	\expval{m}_\infty &= 0,\\
	\langle\!\langle m^2\rangle\!\rangle_\infty &=
    {\varsigma^2\qty(1- \frac{2}{N}) -
    \frac{\varsigma}{\sqrt{e}} + \frac{1}{2} + \mathcal{O}\qty(\nicefrac{1}{N})}.\label{eq:sc_var_reduction_Ramsey}
\end{align}

{
Equation~\eqref{sc-var-ss} allows to derive strict conditions on the initial variance so that in the long-time limit the variance reduces. Fixing an initial state $\rho_m = \mathcal{N}(0, \varsigma) \dyad{\Psi_m}$, we obtain for a given $\ket{\Psi_m}$ the condition $\varsigma>\varsigma^*$, where $\varsigma^*$ is the threshold for narrowing. 
In Table~\ref{t:sc-comparison} we compare $\varsigma^*$ from the uncorrelated protocol in Section~\ref{ss:cooling} and $\Theta$-- and Ramsey--correlated protocols. 
 Furthermore, they also show that achieving a deterministic state, $P_m = \delta_{m,m_0}$ is not possible.
Appendix~\ref{a:bounds-variance} presents derivations and finite--$N$ results.
}

\begin{table}
{
\centering
\begin{tabular}{c| c | c | c}
    $\ket{\Psi_m}$ &  $k$ & $\varsigma^*(N\to\infty)$ &$\varsigma^*(N)$\\
    \hline
    $ \ket{\downarrow}$ & 0 & $\sqrt{\nicefrac{N}{8}} \approx 11.12$ & $\approx 11.12$\\
    $|\Psi^\Theta_m\rangle$  & $\sqrt{2\pi}^{-1}$ & $\sqrt{\nicefrac{\pi}{8}} \approx 0.627$ & $\approx 0.626$ \\
    $|\Psi^R_m\rangle$ & $\sqrt{4e}^{-1}$  & $\sqrt{\nicefrac{e}{4}} \approx 0.824$ & $\approx 0.823$
    \end{tabular} 
    \caption{For an initial state $\rho_m = \mathcal{N}(0, \varsigma) \dyad{\Psi_m}$, and $\langle \!\langle S_z m \rangle \!\rangle_0  = -k \varsigma $, variance reduction can only happen if $\varsigma > \varsigma^*$. 
     Setting $N=1000$, we confirm the efficiency of anti-correlated states in reducing the variance, and contrast leading order corrections of $\varsigma^*$ in $\nicefrac{1}{N}$ with exact results for finite $N$.     Expressions for the latter in App.~\ref{a:bounds-variance}. }
    \label{t:sc-comparison} 
}
\end{table}

Beyond narrowing, the Ramsey correlated state allows the formation of satelite peaks: As a function of $m$, the Bloch vector of the state rotates in the $xz-$plane, creating points where the probability distributions $P_m$ accumulate. Here, since $\hat{B}_m = \hat{e}_z$, the angle between $\vec{B}_m$
and the Bloch vector of the initial state
is simply $\theta_m $.
This is illustrated in Fig.~\ref{fig:sc-purification}. We note that a more slowly rotating field $\theta_m$ (i.e., smaller $\alpha$) results in a narrowing of the distribution as the satellite peaks are largely offset. 

Below, we discuss how states of the form of Eq.~\eqref{ramsey-state-sc} can be prepared in semiconducting quantum dots using Ramsey interferometry, building on Ref.~\cite{Jackson2022}. In superconducting qubits this preparation is challenging due to the ubiquitous dissipative processes, and may require more elaborate protocols and decoupling techniques~\cite{Cywinski2008}.

\section{Quantum-dot spin qubit}\label{s:sq}
We use the MARE to understand the working principles behind a series of recent experiments in which a spin qubit is used to manipulate its nuclear environment~\cite{Jackson2022, Nguyen2023, Hogg2024}. These systems exhibit a wealth of interesting features at different time scales, due to the prominent role of strong system-bath interactions and system-bath correlations. Despite the complexity of these systems, we construct results based on the analytic solution of Eq.~\eqref{genMARE}. 
Beyond reproducing the qualitative aspects of these experiments, the MARE provides a landscape of possibilities in engineering out-of-equilibrium nuclear states. 

\subsection{System and model}
In a semiconducting quantum dot, the Zeeman-split ground state of an electron or hole can be used to implement a spin qubit. The Zeeman field also affects the nuclear spins of the semiconductor substrate which lie in the range of the spin qubit's wave-function~\cite{Coish2004, Klauser2006} and couple to the spin qubit via  hyperfine interactions. The
nuclei have sharp frequencies, associated to the chemical
elements in the host semiconductor~\cite{Crywinsky2009}. For instance, in GaAs quantum
dots there are  two species, each with $s=\nicefrac{3}{2}$ ~\cite{Nguyen2023}, in InGaAs  ~\cite{Hogg2024} three species with spins-$\nicefrac{3}{2}$ and $\nicefrac{9}{2}$,  both platforms with $N
\approx10^4-10^5$ nuclei, and in Si the dominant isotope has spin$-\nicefrac{1}{2}$~\cite{Cai2025}  with $N\approx 10^3-10^4$ nuclei. 

The interaction between the qubit and the nuclei can have many different terms~\cite{Urbaszek2013, Coish2009}, but Eq.~\eqref{eq:hamiltonian} is general enough to capture many of them. We also consider a time-dependent drive,
\begin{align}
    \vec{B}_0(t) = \omega_S \hat{e}_z + \Omega\cos(\omega_d t) \hat{e}_x.
\end{align}
We are particularly interested in the case in which the Zeeman field induces a large splitting in the system~\cite{Coish2008decay}, i.e. $\omega_S \gg \omega_k,\Omega$. In this limit, many terms in Eq.~\eqref{eq:hamiltonian} vanish under the rotating wave approximation and we obtain~\cite{Ribeiro2015, Jackson2022, appel2024, Shofer2025}
\begin{align}
\label{hamqd}
    H =& \vec{B}_0 \cdot \vec{S} + \sum_{k=1}^N  \omega_k I_z^k\\
    &+  S_z\qty[ A_\text{c} J_z + A_\text{nc} (J_x + J_y)],  \nonumber\\
    \vec{B}_0 =& \Delta \hat{e}_z + \Omega \hat{e}_x.
\end{align}
where we wrote the Hamiltonian in the rotating frame relative to the drive frequency. We assume the frequencies $\omega_k$ to be distributed around $\omega_B$ with a Lorentzian with width $\gamma$. We note that other decoherence mechanisms, such as spin diffusion, may alternatively be invoked to justify the MARE~\cite{Paget1977, Paget1982, Coish2008decay, Wust2016, Jackson2022, Zaporski2023}.

The noncolinear terms, which are responsible for changing the magnetisation of the nuclear spin bath are small in comparison to the colinear term~\cite{Ribeiro2015, Shofer2025}.
Indeed, the typical values of $A_\text{c}$ do not allow for treating the colinear term perturbatively~\cite{Klauser2006, Coish2008decay, Hogele2012, Sarma2012}. This leads to an autonomous feedback mechanism between the system and bath in which the system polarises the bath (Knight field)
and the bath polarises the system (Overhauser field)~\cite{Maletinsky2006,
Stepanenko2006, Klauser2006, Urbaszek2013}.

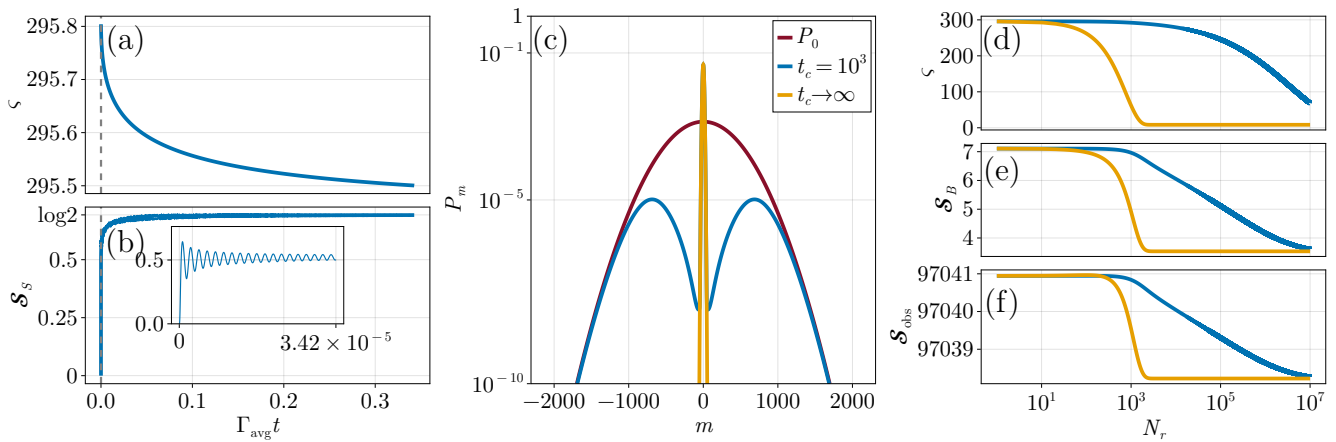
\begin{figure*}[t] 
\centering
\begin{tikzpicture}
\node (a)  at (0,0)
{\includegraphics[width=\linewidth]{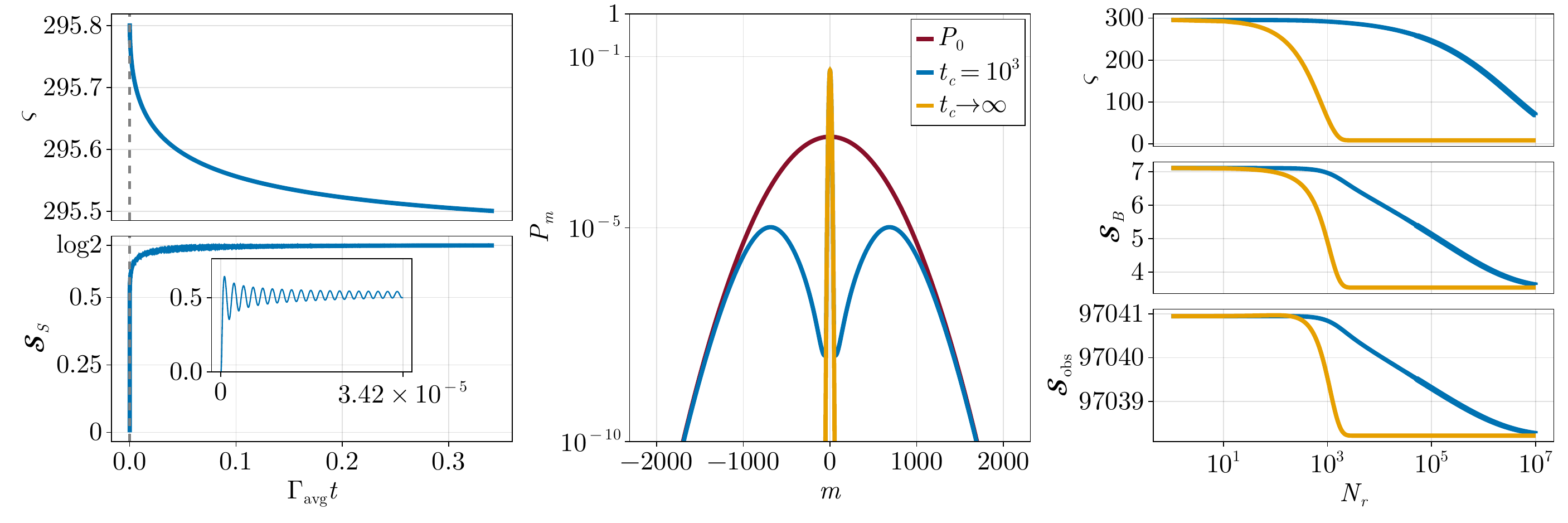}};
\node (a) at (-6.75,2.5) {\large(a)};
\node (b) at (-6.85,-0.2 ) {\large(b)};
\node (c) at (-1.25 ,2.5) {\large(c)};
\node (d) at (4.25, 2.25) {\large(d)};
\node (e) at (4.25, 0.5) {\large(e)};
\node (f) at (4.25, -1.25) {\large(f)};
\end{tikzpicture}
\caption{Narrowing
of nuclear-bath with uncorrelated states. We use nominal values for the electron spin of a GaAs quantum-dot from \cite{Nguyen2023}, listed in Table~\ref{t:sq-exp-values}.
In (a--d) we simulate a single cycle of driving Rabi oscillations at
Hartmann-Hahn resonance initialising state $\ket{\uparrow_z}$ in the
qubit. 
(a) Slight decrease of the standard deviation during a single interaction cycle.
(b) The von Neumann entropy of the system tends to the maximum value, but exhibits oscillations at short time scales (see inset) due to the development of sizeable system-bath correlations.
In (c--f) we
simulate the repeated process of narrowing the bath  up to $N_r=10^7$ repetitions. We consider resetting the spin after the system-bath compound reach the steady state, $\Gamma_\text{avg} t_c \gg 1$ and at finite time, $\Gamma_\text{avg} t_c =3.42\times 10^{-5}$  [vertical-dash in (a/b)], corresponding to $t_c=1000$ ns. 
(c) $P_m$ after narrowing.
(d) Standard deviation at the end
of each cycle.
(e) Sizeable decrease in the Shannon entropy of $P_m$.
(f) The change in observational entropy is dominated by the change in Shannon entropy.
}
\label{fig:sq-panel} 
\end{figure*}

From the Hamiltonian in Eq.~\eqref{hamqd}, we obtain the MARE in Eq.~\eqref{genMARE} with parameters
\begin{align}
\label{eq:sq-mare}
\vec{B}_{m} &= \Omega \hat{e}_x + (\Delta + A_{\rm c}m)\hat{e}_z,\\
\Gamma_m &= 2\pi V_m r_m \varrho(\xi_m + \xi_{m-1})
\qty|\bra{\downarrow_m}S_z\ket{\uparrow_{m-1}}|^2,
\\
\varrho(\omega) &= 4 A_\text{nc}^2 N \frac{\gamma}{(\omega_B - \omega)^2 + \gamma^2}, \\
r_m &= \frac{2}{3}s(s+1) + \frac{m}{N},
\end{align}
where $ \xi_m = \sqrt{\Omega^2+(\Delta+mA_{\rm c})^2}/2$, and $\varrho(\omega)$ is the nuclear-bath spectral density.
The volume factors $V_m$ depend on the type of nuclear spins. For spin-$\nicefrac{1}{2}$, they are given in Eq.~\eqref{vm}. For a bath of spin-$s$, we use the
large number of nuclei spins and apply the central limit theorem to show that (see App.~\ref{a:sspin})
\begin{align}
V_m \approx \frac{[\nicefrac{4}{3} N s (s+1)]!}{[\nicefrac{2}{3} N s (s+1)+m]! [\nicefrac{2}{3} N s (s+1)-m]!},
\end{align} 
with $m\in[-s N, s N]$.  We note that in strained quantum dots, quadrupole terms in the interaction can lead to bath transitions $m\to m\pm 2$ in the MARE~\cite{Gangloff2019, Coish2009, Urbaszek2013}, and the general form of Eq.~\eqref{genMARE} has to be adapted.

There are two important features that distinguish the MARE for the quantum-dot spin qubit from the superconducting qubit scenario discussed above. First, the field $\vec{B}_m$ rotates as a function of $m$. The states $\ket{\uparrow_m}$ thus obtain an explicit $m$-dependence with important consequences for active reservoir engineering. Second, the rates $\Gamma_m$ have a strong dependence on $\Omega$, and due to the large values of $A_{\rm c}$, they also strongly depend on $m$. The rates can therefore be switched off, i.e. $\Gamma_m\rightarrow 0$ for $\Omega\rightarrow 0$. This is a consequence of the large mismatch in frequency $\omega_S\gg\omega_B$. In the absence of a drive, this mismatch prevents energy exchanges between the system and the bath. As we discuss below, this possibility of switching off energy exchanges allows to create system-bath correlations which can then be used for active reservoir engineering. We also note that the $m$-dependence of $\Gamma_m$ implies that the environment can only exchange energy with the system when they are approximately resonant, i.e., $|\vec{B}_m|=2\xi_m\simeq\omega_B$. Once $||\vec{B}_m|-\omega_B|\gg\gamma$, energy exchanges are no longer possible. 

Note that previous theoretical works often rely on mean field type approaches, leading to \textit{classical} rate or Fokker-Plank master equations for the magnetisation distribution~\cite{Danon2008, Hogele2012, Urbaszek2013, Yang2013}. These approaches cannot account for system-bath correlations, which are non-negligible due to the strong colinear term and useful for active reservoir engineering with initial correlations. 
Furthermore, Eq.~\eqref{eq:sq-mare} does not have the same form as the master equations from Refs.~\cite{Riera2021, Riera2022}, due the non-perturbative effect of $A_\text{c}$.

Below, we concentrate on actively engineering a spin-bath of $N=7.0\times
10^4$ spin-$\nicefrac{3}{2}$ ${\ce{_{}^{75}As}}$ nuclei, with an initial high-temperature distribution such that $P_m\propto V_m$, using the parameters from Table~\ref{t:sq-exp-values}. 
\begin{table}[t]  
\centering
\begin{tabular}{c| c}
    Parameter & Value\\
    \hline
    $\omega_{S}$ & $2\pi \times 4.2$~GHz \\
    $\omega_{B}$ & $2\pi  \times 18.96$~ MHz\\
    $A_\text{c}$ & $-2\pi \times 0.13$~MHz\\ $A_\text{nc}$ & $2\pi
    \times 3\times 10^{-3}$~MHz\\ $\gamma$ & $\omega_B/5$\\ $N$ & { $7\times10^4$}\\
    $s$ & $3/2$ \end{tabular} 
    \caption{We use the available reference values of GaAs quantum dot from
    \cite{Nguyen2023, NguyenThesis}. 
    $A_\text{nc}$ is estimated from the $g-$factor anisotropy as suggested in~\cite{appel2024, Shofer2025}, from experimental data~\cite{NguyenThesis}}
    \label{t:sq-exp-values} 
\end{table}

\subsection{Nuclear-bath narrowing without correlations} 
We first consider narrowing the bath distribution by repeatedly initialising the state $\ket{\uparrow_z}$ and driving Rabi oscillations at Hartmann-Hahn resonance, i.e. $\Omega= \omega_B$, as described in Sec.~\ref{s:over} and depicted in Fig.~\ref{fig:overview}. This state results in narrowing because its Bloch vector points in the same (opposite) direction as $\vec{B}_m$ for large negative (positive) values of $m$.
The nonlinearities in $m$ contained in the MARE endow the system with rich and challenging physics. Yet, due to the conserved quantity, the MARE can be analytically solved by the method from Ref.~\cite{Riera2021}, as we explain in App.~\ref{ss:analytic-sol}.

After each initialisation of the system, the variance of $P_m$ and the bath entropy decrease, while the system entropy increases, see Fig.\,\ref{fig:sq-panel}\,(b--d). The relevant timescale for the system-bath interaction is estimated as,
\begin{align}\label{avg-rate}
    \Gamma_\text{avg} = \frac{1}{2sN}\sum_{m=-sN}^{sN}\frac{\Gamma_m}{{V_m}}.
\end{align}
While the quantities of the bath exhibit a monotonic evolution to their steady-state values, the  von Neumann entropy of the system shows an intricate behaviour, as seen in Fig.~\ref{fig:sq-panel} (b). In general, the system density matrix evolves in a non-Markovian fashion~\cite{Breuer2006,Breuer2007, Riera2021}. Compared to the superconducting scenario above, the non-Markovianity is enhanced because the direction of the magnetic field $\vec{B}_m$ depends on the previous energy exchanges between system and bath.

By repeatedly initialising the system in the state $\ket{\uparrow_z}$, significant narrowing can be achieved as illustrated in Fig.\,\ref{fig:sq-panel}\,(d--f). We consider $N_r=10^7$ initialisations, where the system exchanges energy with the environment for the duration of $\Gamma_\text{avg}t_c =3.42 \times 10^{-5}$ (corresponding to $t_c = 1000$ ns). We find that it takes $N_r \approx N$ cycles for a significant reduction of the variance $\langle\!\langle m^2\rangle\!\rangle$ and the bath entropy. 
Although the initial and final distributions are approximately Gaussian, the intermediate steps can contain non-Gaussian features.
In Fig.~\ref{fig:sq-panel} (e) we observe a substantial reduction in the bath Shannon entropy.
The  observational entropy in Fig.~\ref{fig:sq-panel} (f) is dominated by the large average Boltzmann entropy, but its change is dominated by the change in Shannon entropy, which can be seen by comparing with (f). 

Notably, although using $t_c\to\infty$ is apparently more efficient in terms of the number of cycles needed to engineer the reservoir, in practice this might be much less convenient. Indeed, we estimate that the time to reach the steady state is of the order seconds and iterating many repetitions over such long times can be forbidding. That is, in practice, a compromise between number of repetitions and driving time has to be found.

\subsection{Active engineering with correlations}
\label{s:sq_corr}

With the goal of narrowing the bath distribution in mind, we now discuss how to use correlated states.
As for the superconducting qubit case, we first note that the idealised state for narrowing is given by
\begin{align}
\label{eq:idealsq}
	|\Psi^\Theta_m\rangle = \Theta_{m} \ket{\downarrow_m} + (1-\Theta_{m}) \ket{\uparrow_m}.
\end{align}
This state has a Bloch vector that is (anti-)parallel with the field $\vec{B}_m$ for negative (positive) values of $m$. Therefore, it results in an increase in magnetisation for $m<0$ and in a decrease of magnetisation for $m>0$. However, we are not aware of a strategy to prepare this state in an experiment.

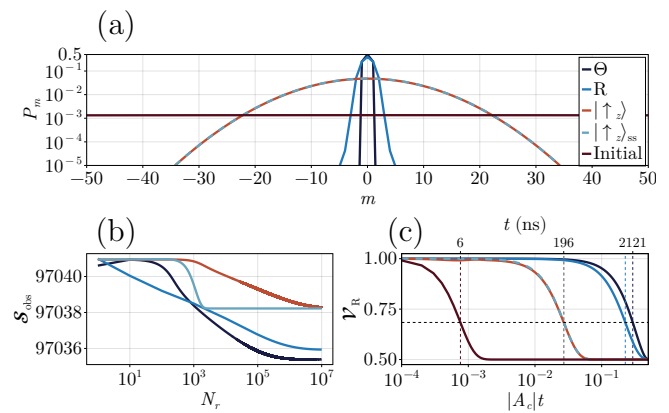
\begin{figure}[t] 
\centering
\begin{tikzpicture}
\node (fig)  at (-6,0) {\includegraphics[width=\linewidth]{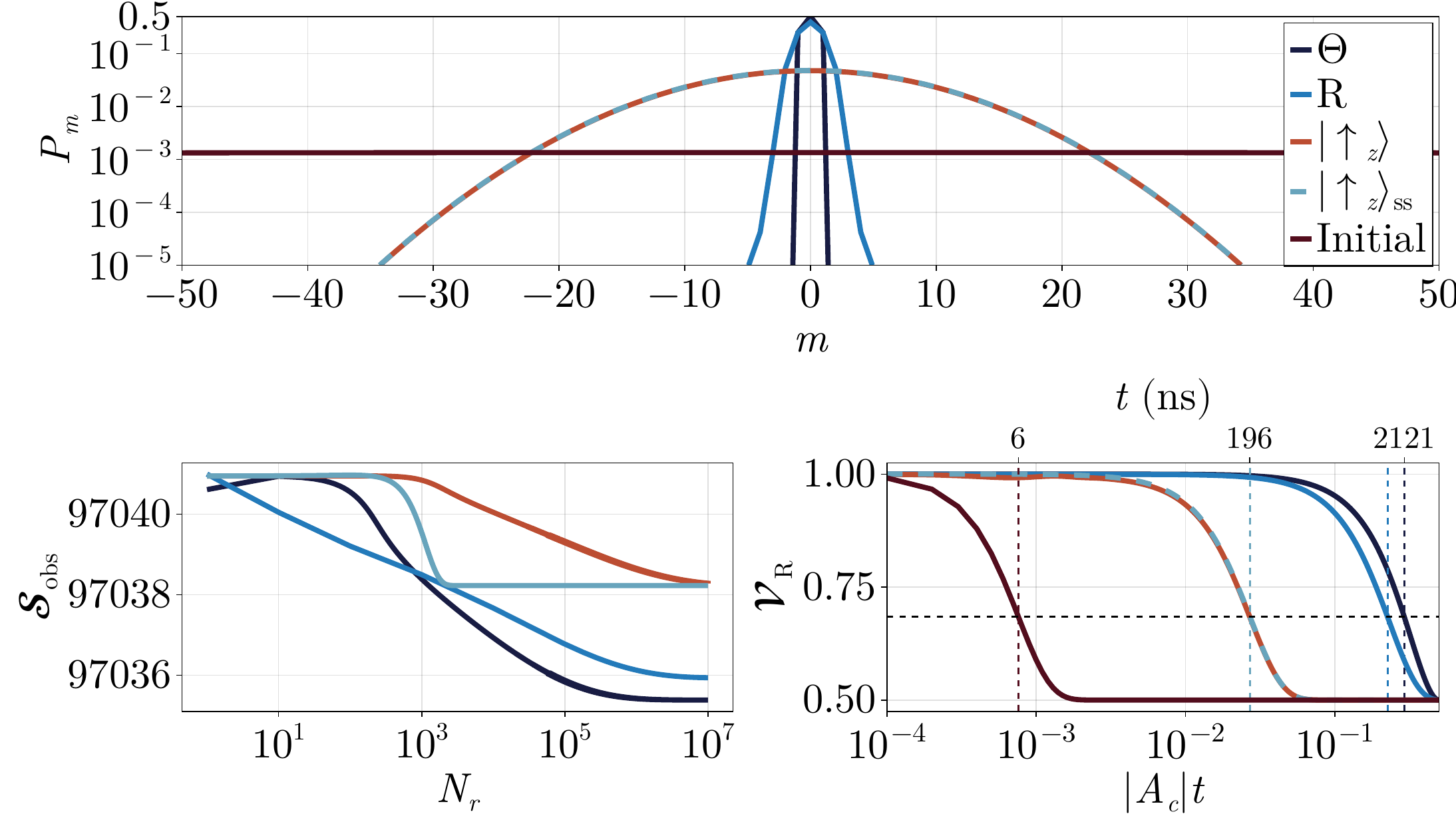}};
\node (a) at (-8.85,2.75) {\large(a)};
\node (b) at (-8.85, 0) {\large(b)};
\node (c) at (-5, 0) {\large(c)};
\end{tikzpicture}
\caption{Engineering narrow nuclear states. We compare distinct strategies using correlated and uncorrelated states to create a concentration of probability at $m=0$. The dark-blue ($\Theta$) represents the ideal protocol, with initial state in Eq.~\eqref{eq:idealsq}. The light blue indicates the Ramsey protocol (R), in which the sensing time is varied between the cycles; for this protocol
$t_c = 10^{4},~ \tau \in[0.02 /|A_{\rm c}|, 0.1 /|A_{\rm c}|]$ with $N_\tau=100$ steps in each run. The overlapping light blue and light red curves correspond to repeated initialisation of $\ket{\uparrow_z}$ for a finite $t_c$ and for $t_c\to\infty$ (subscript ss), see also Fig.~\ref{fig:sq-panel}. Dark red indicates the initial bath distribution. (a) $P_m$ obtained after $N_r=10^7$ repetitions of each protocol. (b) Observational entropy as a function of $N_r$. (c) Ramsey visibility, $\mathcal{V}_\text{R}$, characterising the coherence time, $T_2^*$, of the spin qubit after narrowing.
}
\label{fig:narrowing}
\end{figure}

Therefore, we turn to the quantum sensing protocol introduced in Ref.~\cite{Jackson2022}, which relies on Ramsey interferometry~\cite{Degen2017} to prepare correlated states. According to
Eq.~\eqref{eq:sq-mare}, the dissipative terms are negligible whenever we operate
the qubit far from Hartmann-Hahn resonance; however, the Overhauser field is
always present, and allows the qubit to sense the bath magnetisation. For $\Omega=0,$ the MARE reduces to
\begin{equation}
    \label{eq:MAREsense}
    \partial_t\rho_m = -i [\vec{B}_m\cdot\vec{S},\rho_m].
\end{equation}
Preparing the state $|\psi_m\rangle = (|\!\uparrow_z\rangle+i|\!\downarrow_z\rangle)/\sqrt{2}$, letting it evolve according to Eq.~\eqref{eq:MAREsense}, and rotating the state back to the $xz$-plane (the plane in which $\vec{B}_m$ lies), we obtain the state
\begin{align}
\label{ramseystate}
    |\Psi^R_m\rangle = \cos (A_{\rm c} m
\tau)\ket{\uparrow_z} + \sin(|A_{\rm c}| m
\tau)\ket{\downarrow_z},
\end{align}
where we set $\Delta =0$ and $\tau$ denotes the evolution time. Due to the distribution of the magnetisation, this evolution results in decoherence of the spin qubit $\rho_S=\sum_m P_m|\Psi^R_m\rangle\langle \Psi^R_m|$. By plotting the probability of finding the qubit in the $\ket{\uparrow_z}$ state, the coherence time of the qubit can be determined. This is illustrated in Fig.~\ref{fig:narrowing} (c), and discussed below. For a Gaussian distribution $P_m\approx \mathcal{N}(\mu=0, \varsigma)$, the coherence time is given by $T_2^*=\sqrt{2}/(\varsigma |A_{\rm c}|\tau)$.

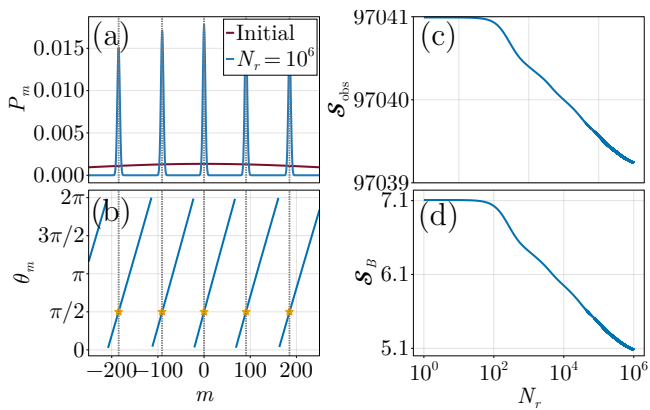
\begin{figure}[t] 
\centering
\begin{tikzpicture}
\node (fig)  at (-6,0) {\includegraphics[width=\linewidth]{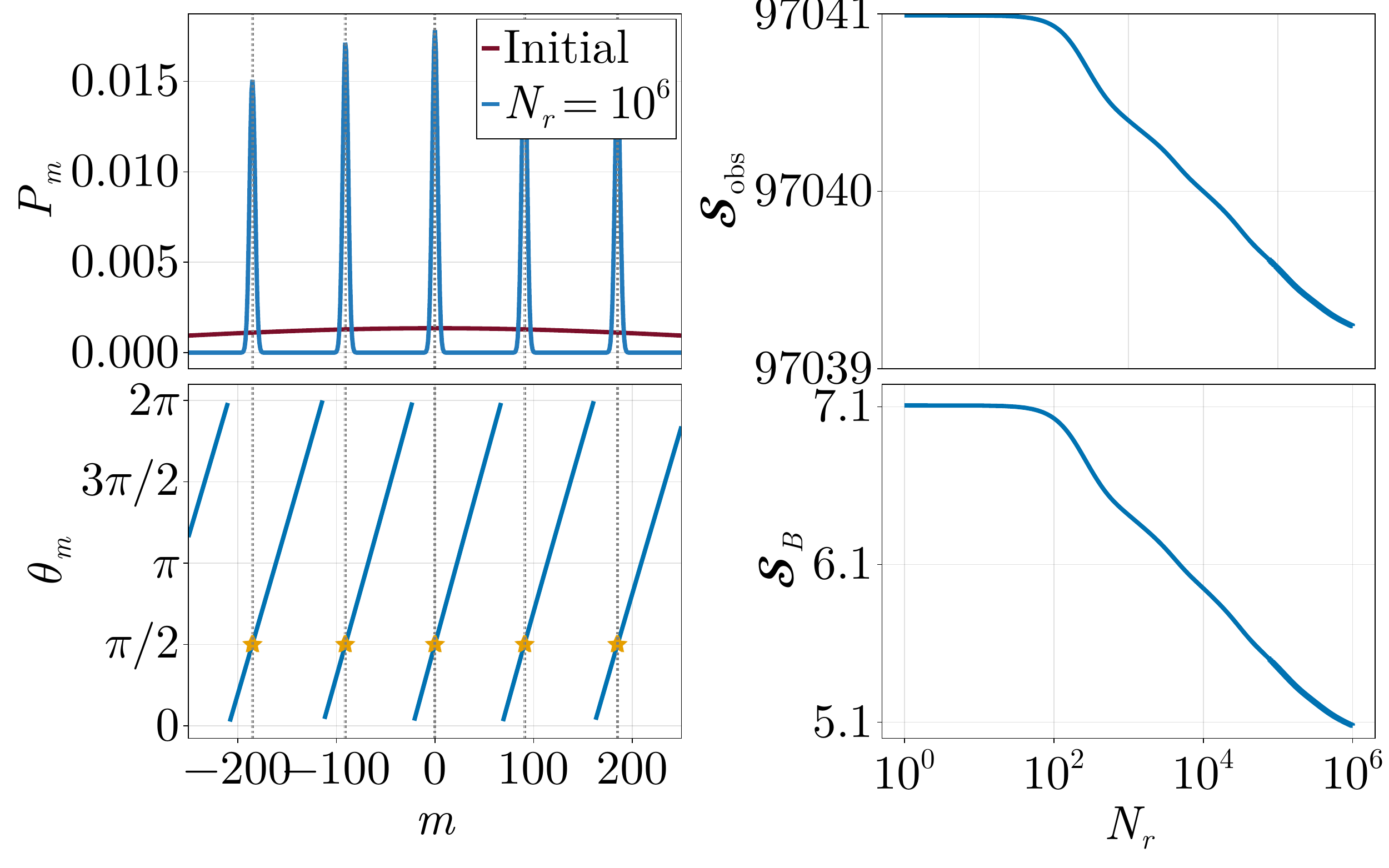}};
\node (a) at (-8.85,2.35) {\large(a)};
\node (b) at (-8.85, 0.0) {\large(b)};
\node (c) at (-4.5, 2.35) {\large(c)};
\node (d) at (-4.5, -0.05) {\large(d)};
\end{tikzpicture}
\caption{Active reservoir engineering with Ramsey-correlated states. Fixed parameters from Table~\ref{t:sq-exp-values} and sensing time, ~$ |A_{\rm c}| \tau = 1/100$.
~(a) Spiked distribution after $10^6$ repetitions and initial distribution.
~(b) $\theta_m$ is the angle between $\vec{B}_m$ and the Bloch vector of $\ket{\Psi^R_m}$. The peaks in (a) are formed every time the angle is $\pi/2$.
~(c) Decrease in observational entropy.
~(d) The decrease in observational entropy is dominated by the decrease in Shannon entropy of the bath. Although many peaks remain, the uncertainty on the value of $m$ reduces.
}
\label{fig:sq-correlated-panel} 
\end{figure}

Here, we consider a repeated initialisation of the state in Eq.~\eqref{ramseystate} with $\tau=(100|A_{\rm c}|)^{-1}$. Since the Bloch vector of this state, as well as $\vec{B}_m$, rotate as a function of $m$, we find multiple regions where the Bloch vector is (anti-)parallel to the field. This results in multiple peaks in $P_m$ when the state is initialised many times, see Fig.~\ref{fig:sq-correlated-panel}. These peaks occur when the Bloch vector of the state is perpendicular to the field $\vec{B}_m$.

Finally, we illustrate how the state in Eq.~\eqref{ramseystate} can be exploited for narrowing $P_m$, as done experimentally in Refs.~\cite{Jackson2022, Nguyen2023, Hogg2024}. To this end, we note that the location of the satellite peaks (at $m\neq0$) depends on the time $\tau$. By sweeping $\tau$, i.e., using a different $\tau_j$ for each initialisation, these satellite peaks can be suppressed, as  in Fig.~\ref{fig:narrowing}. We consider a protocol where $\tau_j$ is increased from $\tau_i = 0.025 |A_{\rm c}|^{-1}$ to $\tau_f = 0.1 |A_{\rm c}|^{-1}$ in $100$ steps. These $100$ preparations are then repeated $10^7$ times.

As seen in Fig.~\ref{fig:narrowing}, this protocol ($R$, light-blue curves) results in  narrowing and a coherence time that is comparable to what is obtained from the ideal state in Eq.~\eqref{eq:idealsq} (dark blue curves). We also compare the results with those obtained from repeated initialisation of uncorrelated states, $\ket{\uparrow_z}$, see also Fig.~\ref{fig:sq-panel}.
The coherence time is determined through the Ramsey visibility~\cite{Degen2017},
\begin{align}
    \mathcal{V}_\text{R} = \frac{1}{2} - \frac{1}{2}\E{\cos(\tau A_{\rm c} m)}.
\end{align}
Where we considered the average with respect to each of the probabilities obtained from different  protocols.
Figure~\ref{fig:narrowing} (c) shows that the ideal state, $\Theta$, results in a value of $T_2^*=2021$\,ns that is only around $ 30\%$ longer than the Ramsey protocol, which gives $T_2^*=1638$\,ns. This is quite remarkable considering that there is no known protocol to prepare Eq.~\eqref{eq:idealsq} whereas the protocol relying on Ramsey interferometry with varying $\tau_j$ has been implemented experimentally~\cite{Jackson2022,Nguyen2023, Hogg2024}. However, the ideal state typically allows for reaching the stationary limit of $N_r \to \infty$ faster, as illustrated in Fig.~\ref{fig:narrowing} (b).
Notably, the correlated states outperform the preparation of $\ket{\uparrow_z}$ by an order of magnitude in the coherence time.

Despite using a simplified model (e.g. a single nuclear species) the MARE predicts similar coherence times to the ones observed experimentally using analogous narrowing protocols in Ref.~\cite{Nguyen2023}.

\subsection{Validity of the assumptions for the spin qubit}\label{ss:sq_validity}
{
To showcase how the validity of the MARE can be assessed in a concrete case, we discuss the assumptions presented in Section~\ref{ss:assumptions} for the quantum-dot spin qubit, using the experimentally relevant parameters listed in Table~ \ref{t:sq-exp-values}.

\textbf{(i)} The relevant bath observable is the collective magnetisation $m$, corresponding to the eigenvalue of $J_z = \sum_k I_k^z$. This choice is physically motivated by the fact that the magnetisation constitutes the dominant source of dephasing for the spin qubit via the  Overhauser field. 

\textbf{(ii)} The interaction term proportional to $A_c J_z$ must be treated non-perturbatively, due the large value of $A_c$. In contrast, the non-collinear coupling $A_\text{nc}$ induces flip-flop processes that change the magnetisation and is weak. It can therefore be treated perturbatively, giving rise to the transition rates $\Gamma_m$ in the MARE. This separation of scales is well satisfied for the parameters in Table~\ref{t:sq-exp-values}, where $A_\text{nc} \ll A_c$.

\textbf{(iii)} The Markov approximation requires that bath correlation functions decay rapidly compared to the system--bath interaction timescale. For the spin qubit, a conservative estimate to assure the validity of the Markov approximation is obtained by comparing relaxation times $\tau_m = (\Gamma_m/V_m)^{-1}$ with the bath correlation decay $\tau_B =\gamma^{-1}$, leading to the conservative condition
\begin{align}
\frac{4\pi}{3}\frac{A_\text{nc}^2}{\gamma^2} N s(s+1) \ll 1,
\end{align}
which is derived in App.~\ref{a:approximations}. Using the parameters in Table~\ref{t:sq-exp-values}, this condition is satisfied with $4\pi A_\text{nc}^2 N s(s+1) /(3\gamma^2) \approx 10^{-1}$.

\textbf{(iv).} The secular approximation requires that the energy splitting induced by the effective field $\vec B_m$ is large compared to the transition rates, $\Gamma_m$. A conservative condition reads
\begin{align}
\frac{2\pi}{3}\frac{A_\text{nc}^2}{\gamma \Omega} N s(s+1) \ll 1,
\end{align}
which is also derived in App.~\ref{a:approximations}. Using the the parameters in Table~\ref{t:sq-exp-values}, the condition is satisfied with
$2\pi A_\text{nc}^2 N s(s+1)/(3\gamma\Omega) \approx 10^{-2}$.

Taken together, these considerations show that the assumptions underlying the MARE are satisfied for the spin-qubit model.
}

\section{Conclusions and outlook}\label{s:conclusion}
We have developed a theoretical framework for active reservoir engineering. Our framework provides an intuitive understanding for how repeated preparation of a controlled quantum system can be used to manipulate its environment and drive it towards a target state. It recovers the qualitative behaviour observed experimentally in different platforms, including superconducting qubits coupled to an environment of TLSs~\cite{Spiecker2024}, as well as spin qubits in quantum dots coupled to nuclear spins~\cite{Jackson2022,Nguyen2023,Hogg2024}. Our framework shows that the dynamics of the system and the reservoir is determined by the effective field felt by the system and the Bloch vector of the prepared state. Knowing the effective field as a function of the environment magnetisation thus allows for developing effective strategies for active reservoir engineering.

Our framework enables the development of novel strategies for active reservoir engineering. Furthermore, it may provide insight into the Hamiltonian underlying an experiment. A promising avenue in this direction is to use the periodic narrowing of the nuclear spin distribution to map out the direction of the effective field as a function of magnetisation 
and the underlying bath spectral density. To make quantitative predictions, the approximations underlying our framework have to be carefully examined for the experiment of interest. In particular, long coherence times in the environment may prevent a classical description of the magnetisation. In this case, the MARE provides a relevant benchmark to determine if the observed behaviour of the reservoir is truly nonclassical and cannot be captured by our approach. In addition, the repeated initialisation of the system may require to investigate timescales shorter than the coherence time of the reservoir. This could be achieved by combining our framework with the coarse-graining approach for Markovian master equations~\cite{schaller_2008}. 
In future work, we will also show how the MARE can be expanded to derive a quantum Fokker-Planck master equation, akin to Ref.~\cite{Annby2022}, for a large bath with short-range jumps.
Another promising avenue is provided by modifying the projector operators, allowing for coherent features in the reservoir~\cite{Breuer2007} to describe dark states~\cite{appel2024, Cai2025}, and include further non-perturbative contributions~\cite{Barnes2011, Tsyplyatyev2011, Wu2012}. The MARE describes a quantum system interacting with a reservoir that decoheres with respect to a given basis; it could be applied to contexts other than active reservoir engineering, such as the emergence of classical outcomes in a detector---treated as a reservoir---coupled to a quantum system~\cite{Allahverdyan2013}.

\begin{acknowledgements}
We acknowledge fruitful discussions with G. N. Nguyen, M. R. Hogg, R. J. Warburton, L. I. Glazman, G. Haack, N. Brunner, N. Gisin, D. Basko, A. Riera-Campeny,  P. Strasberg, and W. Zhang. We also thank an anonymous referee for important comments on our results. This work was supported by the Swiss National Science Foundation (Eccellenza Professorial Fellowship PCEFP2\_194268). 
M.B. acknowledges funding from the European Research Council (ERC) under the European Union’s Horizon 2020 research and innovation programme (Grant agreement No. 101002955 – CONQUER).
\end{acknowledgements}

\section*{Author Contributions}
M.J. carried out the main theoretical development, analytical calculations, numerical simulations, figure preparation, and manuscript writing. M.B. contributed to the theoretical analysis, interpretation of results, and manuscript revision. P.P.P. led and supervised the project, contributed to the theoretical framework and interpretation of the results, and edited the manuscript. All authors discussed the results and approved the final manuscript.

Large language model tools were used to assist with language editing and with the optimisation of simulation code. All scientific content, calculations, simulations, conclusions, and final text were reviewed and approved by the authors, who take full responsibility for the work.
\section*{Data Availability}
All simulations and data generated in for this article is publicly available in the GitHub repository~\cite{Janovitch_MAREPublic2026}.

\bibliography{references}

\onecolumn
\appendix
\section{General Theory}\label{a:deriv}
In this Section we provide a derivation of the MARE for general quantum systems. We discuss important properties and assumptions and provide a general quantum Fokker-Planck master equation via  jump-length expansion.
\subsection{Microscopic derivation of the generalised MARE}
The MARE is a generalisation of the results of Ref.~\cite{Riera2021,Riera2022}, in which we include the part of the system-bath interaction that does not lead to jumps in the bath nonperturbatively.

\textbf{Projectors and rotating frame.---}We consider a system and an environment (bath) with Hamiltonian,
\begin{align}
    H = H_S + H_B + H_{SB}.
\end{align}
We further consider a bath observable with spectral decomposition
\begin{align}
 M_B = \sum_m m \Pi_m,
\end{align}
that commutes with the bath Hamiltonian, $[M_B,H_B] = [\Pi_m,H_B] = 0$. This will be the bath observable that we keep track of. In the main text, this corresponds to the magnetisation and in Refs.~\cite{Riera2021,Riera2022} $M_B$ corresponds to a coarse-grained version of $H_B$. Here we leave it general. The operators $\Pi_m$ are orthogonal projection operators, $\Pi_{m}\Pi_{m'} = \delta_{m,m'}\Pi_m$, but not necessarily rank-1; their rank is denoted $V_m = \tr\Pi_m$ and called \textit{volume factor}. With the help of these quantities, we introduce the correlated projector superoperator,
\begin{align}
    \mathcal{P}\rho = \sum_{m} \rho_m\otimes \frac{\Pi_m}{V_m}, \hspace{2cm} \rho_m = \tr_B\{\Pi_m\rho\}.
\end{align}
Note that when $m$ corresponds to a coarse-grained energy, then the state $\Pi_m/V_m$ denotes the microcanonical ensemble~\cite{Riera2021}.

We now decompose the $H_{SB}$ term in the Hamiltonian in two parts,
\begin{equation}
\label{split}
H_{SB} = \delta H + V,\hspace{2cm}    \delta H = \sum_m \expval{H_{SB}}_m\Pi_m,
\end{equation}
where we introduced the microcanonical average $\expval{\circ}_m = \tr_B\{\circ\Pi_m/V_m\}$. 

We will only treat $V$ perturbatively (not to be confused with the volume factors, $V_m$). The rest of the interaction, $\delta H$, is treated non-perturbatively. To this end, we consider the unitary operator,
\begin{align}
\label{eq:rotation}
    R(t) = e^{it(H_S + H_B + \delta H)} = e^{it(H_S +\delta H)} e^{it H_B},
\end{align}
where we note that, in general $[H_S, \delta H] \neq 0$. We may write
\begin{equation}
\label{eq:rot}
R(t) = \sum_m R_m(t) \Pi_m e^{it H_B},
\end{equation}
with
\begin{align}
    R_m(t) = e^{it H_m},\hspace{2cm}    H_m = H_S + \expval{H_{SB}}_m.
\end{align}
If the system has only two levels, we can always write $H_m = \vec{B}_m\cdot \vec{S}$ where $S_{x,y,z}$ are spin operators---we will not make this assumption during the derivation below.

We now consider the density matrix in the rotating frame
\begin{equation}
    \label{eq:rotatingdens}
    \tilde{\rho}(t) = R(t)\rho(t) R^\dagger(t),\hspace{2cm}\partial_t\tilde{\rho}(t)=\mathcal{V}(t) \tilde{\rho}(t),
\end{equation}
with $\mathcal{V}(t)\circ = -i[\tilde{V}(t), ~\circ]$, and $\tilde{V}(t)=R(t)VR^\dagger(t)$. Using $\mathcal{Q}=1-\mathcal{P}$, one may then formally solve the equation $\partial_t\mathcal{Q}\tilde{\rho}$ and plug the solution into the equation $\partial_t\mathcal{P}\tilde{\rho}$. Using $\mathcal{P}\mathcal{V}\mathcal{P} = 0$, this results in the standard Nakajima-Zwanzig equation \cite{Nakajima:1958,Zwanzig:1960},
\begin{align}
    \partial_t\mathcal{P}\tilde{\rho} =  \int_0^{t}\dd t \mathcal{P}\mathcal{V}(t)\mathcal{G}(t,t')\mathcal{Q}\mathcal{V}(t'){\mathcal{P}}\tilde{\rho}(t'),    
\end{align}
where {$\mathcal{G}(t,t')=\mathcal{T}\exp{\int_{t'}^t\dd s \mathcal{Q}\mathcal{V}(s)}$} is the memory kernel and $\mathcal{T}$ denotes time ordering.

\textbf{Weak coupling.---} We now perturbatively expand the Nakajima-Zwanzig equation in $V$. The lowest non-zero order gives,
\begin{align}\label{microNZeq}
    \partial_t\tilde{\rho}_m  =
    -\sum_{m'}\int_0^t \dd t'\tr_B\Bigg\{\Pi_{m}\qty[\tilde{V}(t),\qty[\tilde{V}(t'), \tilde{\rho}_{m'}(t')\otimes\frac{\Pi_{m'}}{V_{m'}}]]\Bigg\}.
\end{align}
It is always possible to write the interaction in the form 
\begin{align}
V = \sum_\alpha \mathsf{S}_\alpha \otimes \textsf{B}_\alpha,\label{interaction-form}
\end{align}
where $\textsf{S}_\alpha$ acts on the system and $\textsf{B}_\alpha$ on the bath. We then obtain,
\begin{align}
    \tilde{V}(t) =  \sum_{\alpha,m,m'} \tilde{\textsf{S}}_{mm'}^\alpha \otimes  \Pi_m \tilde{\bop}_\alpha\Pi_{m'},\label{vtilde}
\end{align}
where we introduced $\tilde{\bop}_\alpha(t) =e^{it H_B}\bop_\alpha e^{-itH_B}, ~\tilde{\textsf{S}}^\alpha_{mm'}=R_m(t) \textsf{S}_\alpha R^\dagger_{m'}(t)$. Note that, although $R(t)$ in general does not preserve the original product structure of the interaction, mixing operators from system  and bath, the product can be restored at the cost of introducing the sum over $m,m'$. This is the immediate consequence of treating $\expval{H_{SB}}_m$ exactly, which imprints a dependence on $m$ on the transitions in the system induced by the operators $\tilde{\textsf{S}}_{mm'}^\alpha$. Substituting Eq.~\eqref{vtilde} in Eq.~\eqref{microNZeq} we obtain,
\begin{align}
\label{eq:redfield}
   \partial_t \tilde{\rho}_m = \sum_{m'\alpha\beta} \int_0^{t} \dd t' C_{mm'}^{\alpha\beta}(t'-t)
    \qty[\tilde{\textsf{S}}_{mm'}^\alpha(t) \frac{\tilde{\rho}_{m'}(t')}{V_{m'}} \tilde{\textsf{S}}_{m'm}^{\beta}(t')  - \tilde{\textsf{S}}_{mm'}^\alpha(t)  \tilde{\textsf{S}}_{m'm}^{\beta}(t')  \frac{\tilde{\rho}_{m}(t')}{V_{m}}]+ \text{h.c.}
\end{align}
where we introduced the microcanonical correlation functions,
\begin{align}
    C_{mm'}^{\alpha \beta}(t-t') &= \tr_B\qty{ \Pi_{m} \tilde{\bop}_\alpha(t) \Pi_{m'} \tilde{\bop}_\beta(t')}= \tr_B\qty{ \Pi_{m} \tilde{\bop}_\alpha(t-t') \Pi_{m'} \bop_\beta(0)}.\nonumber
\end{align}
The dependence on $t-t'$ instead of $t,t'$ is a consequence of $[\Pi_m, H_B]=0$. The correlation functions have the property,
\begin{align}\label{corr-property}
    C_{mm'}^{\alpha \beta}(t-t') &= C_{m'm}^{\beta\alpha}[-(t-t')] .
\end{align}
The next steps follow closely the microscopic derivation of a Lindblad master equation. We will perform the Markov and secular approximations \cite{breuer:book}. 

\textbf{Markov approximation.---} Assuming that the microcanonical correlation functions decay on a timescale that is much shorter than the characteristic time over which $\tilde{\rho}_m$ changes, we can substitute $\tilde\rho_m(t')$ with $\tilde\rho_m(t)$ in Eq.~\eqref{eq:redfield}. Changing variables to $\tau = t-t'$, we find
\begin{align}\label{gen-markov}
    &\partial_t \tilde\rho_m \approx \sum_{m'\alpha\beta} \int_0^{\infty} \dd \tau C_{mm'}^{\alpha\beta}({-\tau}) 
    \qty[\tilde{\textsf{S}}_{mm'}^\alpha(t) \frac{\tilde{\rho}_{m'}(t)}{V_{m'}} \tilde{\textsf{S}}_{m'm}^{\beta}(t-\tau)  - \tilde{\textsf{S}}_{mm'}^\alpha(t)  \tilde{\textsf{S}}_{m'm}^{\beta}(t-\tau)  \frac{\tilde{\rho}_{m}(t)}{V_{m}}]\nonumber + \text{h.c.}
\end{align}

The integrals can be evaluated by introducing the Fourier modes of the operators
\begin{align}
    \tilde{\textsf{S}}^\alpha_{mm'}(t) = \sum_\omega e^{-i\omega t} \textsf{L}_{mm'}^\alpha(\omega).
\end{align}
As we now show, the frequencies $\omega$ are related to the eigenfrequencies of $H_m$. Introducing the notation, $H_m\ket{\lambda_m} = \xi_{\lambda_m} \ket{\lambda_m}$ we write
\begin{align}\label{mrotation}
    R_m(t) = \sum_{\lambda} e^{it\xi_{\lambda_m}} \dyad{\lambda_m},
\end{align}
resulting in,
\begin{align}\label{stildes}
    \tilde{\textsf{S}}^\alpha_{mm'}(t) = \sum_{\lambda, \nu} e^{it\qty(\xi_{\lambda_m} -\xi_{\nu_{m'}})} F^{\alpha\lambda\nu}_{mm'} \dyad{\lambda_m}{\nu_{m'}},
\end{align}
where $F^{\alpha\lambda \nu}_{mm'} = \bra{\lambda_{m}}\textsf{S}_\alpha \ket{\nu_{m'}}$ are complex scalars satisfying,
\begin{align}\label{F-property}
    F_{mm'}^{\alpha\lambda\nu} = (F_{m'm}^{\alpha\nu\lambda})^*.
\end{align}
 In general, the eigenvectors from different $m$'s  are not orthogonal, i.e., $\bra{\nu_{m'}}\ket{\lambda_m}\neq\delta_{\lambda,\nu},$ for $m\neq m'$ . 

Equation~\eqref{stildes} leads to the eigenfrequencies $\omega=\omega(\lambda_m,\nu_{m'}) = - \xi_{\lambda_m} + \xi_{\nu_{m'}}$, and the corresponding Fourier modes, $\textsf{L}^\alpha_{mm'}(\omega)=F^\alpha_{\lambda_{m},\nu_{m'}} \dyad{\lambda_m}{\nu_{m'}}$. Here we consider the case of nondegenerate frequencies, i.e., if $\omega(\lambda_m,\nu_{m'})=\omega(\lambda'_m,\nu'_{m'})$, then $\lambda_m=\lambda_m'$ and $\nu_{m'}=\nu_{m'}'$. To simplify the notation, we carry on the calculation with the summation over $\omega$, keeping in mind the explicit dependence on the eigenvalues of $H_{m}$ and $H_{m'}$. 

Using the Fourier decomposition we find,
\begin{align}\label{pre-secular}
    \partial_t \tilde\rho_m = \sum_{m'\alpha\beta}\sum_{\omega,\omega'}e^{-it(\omega'+\omega)}
    W_{mm'}^{\alpha\beta}(\omega)
    &\qty[\textsf{L}_{mm'}^\alpha(\omega) \frac{\tilde{\rho}_{m'}}{V_{m'}}\textsf{L}_{m'm}^{\beta}(\omega')  - \textsf{L}_{mm'}^\alpha(\omega)  \textsf{L}_{m'm}^{\beta}(\omega')  \frac{\tilde{\rho}_{m}}{V_{m}}]+ \text{h.c.},
\end{align}
where we introduced,
\begin{align}\label{Wrates}
    W^{\alpha \beta}_{mm'}(\omega) ={\int_0^\infty \dd\tau e^{+i\tau \omega} C^{\alpha\beta}_{mm'}(-\tau)} = \int_0^\infty \dd\tau e^{-i\tau \omega} C^{\alpha\beta}_{mm'}(\tau).
\end{align}

\textbf{Secular approximation.---}We retain only terms with $\omega'=-\omega$. However, note that, since $V$ is Hermitian,
\begin{align}
    \sum_{\alpha} \textsf{S}_\alpha\otimes \bop_\alpha = \sum_{\alpha} \textsf{S}_\alpha^\dagger\otimes \bop_\alpha^\dagger.
\end{align}
That is, either the terms in the sum are Hermitian or the sum contains their Hermitian conjugate. For simplicity, let us assume henceforth that at least the system operators are Hermitian, $\textsf{S}_\alpha = \textsf{S}_\alpha^\dagger,~\forall\,\alpha$. This implies that,
\begin{align}
       \textsf{L}_{m',m}^\alpha(-\omega) = [\textsf{L}_{m,m'}^\alpha(\omega)]^\dagger = (F^{\alpha\lambda \nu}_{mm'})^* \dyad{\nu_{m'}}{\lambda_m}.
\end{align}
Rotating back to the original frame, we find the general form of the MARE,
\begin{align}\label{generalMARE}
      &\partial_t \rho_m =-i[H_m + H^\text{LS}_m,\rho_m]+\sum_{m'}\sum_{\lambda \nu}
    \Gamma_{mm'}^{ \lambda \nu}
    \qty[\dyad{\lambda_m}{\nu_{m'}} \frac{\rho_{m'}}{V_{m'}}\dyad{\nu_{m'}}{\lambda_{m}}  -\frac{1}{2}\qty{\dyad{\lambda_m}{\lambda_m},  \frac{\rho_{m}}{V_{m}}}],
    \end{align}
    where
    \begin{align}
    {
    H^\text{LS}_m = \frac{1}{V_m}\sum_{m'}\sum_{\lambda\nu} \zeta^{\lambda \nu}_{mm'}\dyad{\lambda_{m}}\label{eq:lamb}, 
    }
\end{align}
and 
\begin{align}\label{rates}
    \sum_{\alpha\beta}W^{\alpha \beta}_{mm'}F^{\alpha\lambda \nu}_{mm'} \qty(F^{\beta\lambda \nu}_{mm'})^* =:  \Gamma_{mm'}^{\lambda \nu }/2 +  i\zeta_{mm'}^{ \lambda \nu}.
\end{align}
From the properties in Eq.~(\ref{corr-property},~\ref{F-property}) the rates are symmetric tensors, 
\begin{align}\label{gen-balance}
\Gamma_{mm'}^{\lambda\nu} = \Gamma_{m'm}^{\nu\lambda},
\end{align}
which is understood as a generalised detailed balance condition. In general, the above MARE does not have a conserved quantity as Eq.~\eqref{genMARE}. 

Reference~\cite{Breuer2007} is the first occurence of the general form of Eq.~\eqref{generalMARE}, where it is shown  that under very general conditions the generator of non-Markovian evolution of an open quantum system can be written in terms of Eq.~\eqref{generalMARE}, as $\partial_t \rho_S = \sum_m \partial_t\rho_m$. In Ref.~\cite{Breuer2007}, however, no microscopic derivation is provided. Recently, Refs.~\cite{Riera2021, Riera2022} presented microscopic derivations which lead to a particular case of Eq.~\eqref{generalMARE} by treating the interaction entirely as a perturbation. Our method generalises this approach, obtaining the generator of non-Markovian evolution in  its general form, i.e., with $m$-dependent jump operators stemming from non-perturbative contributions of the interaction.

\subsection{Trace-preservation}
The results from Ref.~\cite{Breuer2007} assure that Eq.~\eqref{generalMARE} generates a completely-positive and trace-preserving evolution (CPTP) of $\rho_S$.
Yet, it is instructive to show trace-preservation. Introducing,
\begin{align}
    q(m, \lambda, \nu) &= \bra{\lambda_m} \rho_m \ket{\nu_m} = [q(m, \nu, \lambda)]^*,~\lambda\neq\nu,\\
    p(m, \lambda)& = \bra{\lambda_m} \rho_m \ket{\lambda_m},
\end{align}
we can decouple coherences from populations in Eq.~\eqref{generalMARE},
\begin{align}
    &\partial_t q(m, \lambda, \nu)  =-\Bigg[ i\qty(\bar{\xi}_{\lambda_m} - \bar{\xi}_{\nu_m})+\sum_{m'\eta}\frac{\Gamma_{mm'}^{\lambda \eta} + \Gamma_{mm'}^{\nu \eta}}{2V_m}\Bigg]q(m, \lambda, \nu) \\
    &\partial_t p(m, \lambda)  = \sum_{m'}\sum_{\nu} \Gamma_{mm'}^{\lambda\nu}\qty[\frac{p(m', \nu)}{V_{m'}} - \frac{p(m,\lambda)}{V_m}],\label{pop-equation}
\end{align}
with $(H_m+H^{\rm LS}_m)\ket{\lambda_m}=\bar{\xi}_{\lambda_m}\ket{\lambda_m}$.
Trace preservation of the system density matrix is equivalent to $\sum_{\lambda m} \partial_t p(m,\lambda)=0$. Performing the sum, we conclude that this is the case due to the detailed balance condition in Eq.~\eqref{gen-balance}.

\section{Inhomogeneous Anisotropic Central Spin Model}\label{a:central-spin-mare}
We now particularise the microscopic derivation of App.~\ref{a:deriv} to a Hamiltonian that is more general than Eq.~\eqref{eq:hamiltonian} in the main text, including inohomogeneity in the anisotropy tensor. 
The interaction term can be writen as,
\begin{align}
    H_{SB} = \sum_k\sum_{\alpha = x,y,z}\qty[ A_k^{\alpha, z} S_\alpha I_k^z + S_\alpha(A_k^{\alpha,+}I_k^+ +A_k^{\alpha,-} I_k^-)]
    = \sum_\alpha S_\alpha (\bop^z_\alpha + \bop^+_\alpha + \bop^-_\alpha)\nonumber
\end{align}
where $S_\alpha$ are spin-$\nicefrac{1}{2}$ operators, $A_k^{\alpha,\beta},~\alpha,\beta = x,y,z$ are elements of the inhomogeneous anisotropy tensor, $\bm{A}_k$, and we introduced $A_k^{\alpha,\pm} = A_k^{\alpha, x} \pm i A_k^{\alpha, y}$. In the second equality we introduced the operators,
\begin{align}
    {\sf B}_\alpha^{l} = \sum_k A_k^{\alpha, l} I^{l}_k.
\end{align}

We now split the interaction as in Eq.~\eqref{split}, by noting that
\begin{align}
   \expval{H_{SB}}_m = \sum_\alpha S_\alpha\langle \bop^z_\alpha\rangle_m = m \sum_\alpha \bar{A}_{\alpha,z} S_\alpha, 
\end{align}
with $\bar{A}_{\alpha, z} =\sum_k A^{\alpha,z}_k/N$. Thus
\begin{align}
    \delta H = \sum_\alpha \bar{A}_{\alpha,z}S_\alpha J_z,\hspace{2cm}
    V = \sum_\alpha S_\alpha (\delta\bop^z_\alpha + \bop^+_\alpha + \bop^-_\alpha) = \sum_\alpha S_\alpha \otimes {\sf C}_\alpha,
\end{align}
where $J_z=\sum_k I_k^z$,
\begin{align}
    \delta\bop^z_\alpha  = \sum_k \delta A^{\alpha,z}_k I_k^z,\hspace{2cm}
    {\sf C}_\alpha = \delta\bop^z_\alpha + \bop^+_\alpha + \bop^-_\alpha,
\end{align}
and $\delta A^{\alpha,z}_k = A^{\alpha,z}_k -  \bar{A}_{\alpha,z} $ quantifies the inohomogeneity.
Note that $V$ is now in the form required in the general derivation, Eq.~\eqref{interaction-form}, and the operators involved in the product structure are Hermitian, $S_\alpha^\dagger = S_\alpha,~{\sf C}_\alpha^\dagger= {\sf C}_\alpha$. According to $\delta H$, the $m$-resolved Hamiltonian now reads,
\begin{align}
\label{eq:hamm}
    H_m =\vec{B}_0\cdot \vec{S} + m\sum_\alpha  \bar{A}_{\alpha, z} S_\alpha= \vec{B}_m\cdot \vec{S},\hspace{2cm}
        \vec{B}_m = \vec{B}_0 + m \sum_\alpha \bar{A}_{\alpha,z} \hat{e}_\alpha,
\end{align}
with eigenvalues $\pm \xi_m$ and eigenvectors $\ket{\uparrow_m},~\ket{\downarrow_m}$.
From Eq.~\eqref{eq:rot}, we obtain the unitary transformation,
\begin{align}
    R(t) = {\sum_m e^{it \sum_k \omega_k I_k^z}R_m(t)\Pi_m},\hspace{2cm}
    R_m(t) = e^{i\xi_m t}\dyad{\uparrow_m} + e^{-i\xi_m t}\dyad{\downarrow_m}.\label{Rm}
\end{align}

All the steps of the general derivation now follow. We now seek to achieve the form presented in Eq.~\eqref{genMARE} by computing the correlation functions explicitly. We write,
\begin{align}
    C^{\alpha \beta}_{mm'}(\tau) &= \tr{\Pi_m \tilde{{\sf C}}_\alpha(\tau)\Pi_{m'}{\sf C}_\beta}
    = \sum_{j=+,-,z}C^{\alpha \beta j}_{mm'}(\tau),
\end{align}
with,
\begin{align}
    C^{\alpha\beta,z}_{mm'}(\tau) &= \tr{\Pi_m \delta\tilde{\bop}_\alpha^z (\tau)\Pi_{m'} \delta\bop_\beta^z}\label{corr-deph},\\
    C^{\alpha\beta,+}_{mm'}(\tau) &= \tr{\Pi_m \tilde{\bop}_\alpha^+(\tau) \Pi_{m'} \bop_\beta^-}\label{corr-plus},\\
    C^{\alpha\beta,-}_{mm'}(\tau) &= \tr{\Pi_m \tilde{\bop}_\alpha^-(\tau) \Pi_{m'} \bop_\beta^{+}}\label{corr-minus}.
\end{align}

We start by discussing Eq.~\eqref{corr-deph} which entails processes induced purely by the inohomogeneity.
Since $[\delta B_\alpha^z, I_k^z]=0$, this correlation function is proportional to $\delta_{mm'}$ and is time-\textit{independent}, $\delta \tilde{B}^z_\alpha(\tau)=\delta B^z_\alpha$. The former means that no jumps in the bath are induced and the latter that the correlation function does not decay, violating the Markov approximation. 

{
For the remaining of this derivation, we now focus on an environment composed of spin-$\nicefrac{1}{2}$ constituents. Appendix~\ref{a:sspin} shows how to obtain approximate correlation functions for arbitrary spin operators. 

To evaluate Eq.~\eqref{corr-deph}, we use $[I_k^z,\Pi_m]=0$ which results in
\begin{equation}
\label{eq:pikk}
    \tr{\Pi_m I_k^z\Pi_{m'} I_k^z} = \delta_{m,m'}\frac{V_m}{4},
\end{equation}
when the spin operators act on the same spin.
When the spin operators act on different spins, we may write
\begin{equation}
\label{eq:ikikp}
  \tr{\Pi_m I_k^z I_{k'}^z} =\frac{N^\uparrow(m,N)}{4}\left[p^\uparrow(m,N|\uparrow)-p^\downarrow(m,N|\uparrow)\right]  - \frac{N^\downarrow(m,N)}{4}\left[p^\uparrow(m,N|\downarrow)-p^\downarrow(m,N|\downarrow)\right],
\end{equation}
where $k\neq k'$. Here we introduced the quantity $N_\uparrow(m,N)$ which gives the number of states where a given spin (say the $k$-th spin) points up and similarly for $N_\downarrow(m,N)$. Since the $k$-th spin either has to point up or down, we have
\begin{equation}
\label{eq:nsum}
    N_\uparrow(m,N)+N_\downarrow(m,N) = V_m,
\end{equation}
where the number of states with total magnetisation $m$~\cite{Dicke1954, MandelWolf95}, the volume factors, are given by
\begin{align}
    V_m = \frac{N!}{(\nicefrac{N}{2}+m)!(\nicefrac{N}{2}-m)!}.\label{volume-factor}
\end{align}
The average magnetisation of a single spin can be written as
\begin{equation}
\label{eq:ndiff}
    \langle I_k^z\rangle = \frac{m}{N} = \frac{N_\uparrow(m,N)-N_\downarrow(m,N)}{2V_m}.
\end{equation}
Together, Eqs.~\eqref{eq:nsum} and \eqref{eq:ndiff} imply
\begin{equation}
\label{eq:nksig}
    N_\sigma(m,N)=V_m\left(\frac{1}{2}+\sigma\frac{m}{N}\right),
\end{equation}
where $\sigma = \pm$ for $\uparrow/\downarrow$.
Furthermore, in Eq.~\eqref{eq:ikikp}, we introduced the probability for a given spin to point up, if another spin is known to point down, $p_\uparrow(m,N|\downarrow)$, and similarly for other spin configurations. Since one spin is known, these probabilities can be expressed as
\begin{equation}
\label{eq:pksig}
    p_\sigma(m,N|\sigma')=\frac{N_\sigma(m-\sigma'/2,N-1)}{N_\uparrow(m-\sigma'/2,N-1)+N_\downarrow(m-\sigma'/2,N-1)}=\frac{1}{2}+\sigma\frac{m-\sigma'/2}{N-1},
\end{equation}
where the arguments in $N_\sigma$ reflect the fact that by knowing one spin is in the $\sigma'$ state, the remainder is a spin system with $N-1$ spins and magnetisation $m-\sigma'/2$.
With the help of Eqs.~\eqref{eq:pikk}, \eqref{eq:ikikp},\eqref{eq:nksig} and \eqref{eq:pksig}, we find
\begin{equation}
\label{eq:cz0}
    C^{\alpha\beta,z}_{mm'}(\tau) = \delta_{m,m'}\sum_k\delta A_k^{\alpha,z}\delta A_k^{\beta,z}V_m\frac{N^2-4m^2}{4N(N-1)},
\end{equation}
where we further made use of the fact that $\tr{\Pi_m I_k^z I_{k'}^z}$ does not depend on the values of $k$ and $k'$ and that $\sum_{k'\neq k}\delta A_{k'}^\alpha=-\delta A_k^\alpha$.}

The time independence in Eq.~\eqref{eq:cz0} is a consequence of the fact that our Hamiltonian in Eq.~\eqref{eq:hamiltonian} does not include any $T_1$  processes on the bath (e.g. spin-flips/thermalisation). In reality, the correlation function in Eq.~\eqref{corr-deph} decays with the $T_1$ time of the bath (see, for instance Ref.~\cite{Spiecker2024solomon}). We thus treat this correlation function phenomenologically as
\begin{equation}
\label{eq:corrz}
    C^{\alpha,z}_{mm'}(\tau) = e^{-|\tau|/T_1} C^{\alpha,z}_{mm'}(0),
\end{equation}
where $C^{\alpha,z}_{mm'}(0)$ is given by Eq.~\eqref{eq:cz0}.

We now concentrate on evaluating Eq.~\eqref{corr-plus}. Writing out,
\begin{align}
    C^{\alpha\beta, +}_{mm'}(-\tau) = \sum_{k} A_{k}^{\alpha, +} A_{k}^{\beta, -} e^{-i\tau \omega_k}  r^+_{mm'},\hspace{2cm}
    r_{mm'}^+= \tr{\Pi_m I_k^+ \Pi_{m'} I_k^-},
\end{align}
we find
\begin{equation}\label{rate_plus}
    r_{mm'}^+ = \delta_{m',m-1}\tr{\Pi_m I_k^+I_k^-}=V_{m}\delta_{m',m-1}\left(\frac{1}{2}+\frac{m}{N}\right).
\end{equation}
The correlation function then reads, 
\begin{equation}\label{sc-corr-ft} 
C^{\alpha\beta, +}_{mm'}(\tau)  
=  \delta_{m',m-1} V_{m} \frac{\nicefrac{N}{2} + m}{N} \int_{-\infty}^\infty\dd \omega \varrho_{\alpha\beta}(\omega) e^{i\tau \omega}. \nonumber 
\end{equation} 
In  the last equality we introduced the
spectral densities ~\cite{potts2024quantumthermodynamics}
\begin{align}\label{spectral-density}
\varrho_{\alpha\beta}(\omega)=\sum_{k=1}^N
A_k^{\alpha,+}A_k^{\beta,-}\delta(\omega-\omega_k).
\end{align}
Similarly, we find,
\begin{align}
\label{eq:cmin}
    C^{\alpha\beta, -}_{mm'}&= 
  \delta_{m',m+1} V_{m} \frac{\nicefrac{N}{2} - m}{N} \int_{-\infty}^\infty\dd \omega \varrho_{\alpha\beta}(\omega) e^{-i\tau \omega}.  
\end{align}

We now determine the jump operators featuring in the MARE. Equation~\eqref{Rm} fully determines the Fourier components,
\begin{align}
    \tilde{S}^\alpha_{mm'}(t)=R_m(t)S_\alpha R_{m'}^\dagger(t)=\sum_{\omega}e^{-i\omega t}L^\alpha_{mm'}(\omega),
\end{align}
where
\begin{align}
    L^\alpha_{mm'}(\xi_m+\xi_{m'}) &= F^{\alpha,\downarrow,\uparrow}_{mm'} \dyad{\downarrow_m}{\uparrow_{m'}}\label{eigenops1},\\
    L^\alpha_{mm'}(-\xi_m-\xi_{m'}) &=F^{\alpha,\uparrow,\downarrow}_{mm'} \dyad{\uparrow_m}{\downarrow_{m'}}\label{eigenop2},\\
    L^\alpha_{mm'}(\xi_m-\xi_{m'}) &= F^{\alpha,\downarrow,\downarrow}_{mm'} \dyad{\downarrow_m}{\downarrow_{m'}},\\
    L^\alpha_{mm'}(-\xi_m+\xi_{m'}) &=F^{\alpha,\uparrow,\uparrow}_{mm'} \dyad{\uparrow_m}{\uparrow_m'},
\end{align}
and we used the notation $F^{\alpha,\lambda,\nu}_{mm'} = \bra{\lambda_m}S_\alpha\ket{\nu_{m'}}$. Crucially, the above establish the set of eigenfrequencies,
\begin{align}\label{eigenfreqs}
    \{\xi_m + \xi_{m'},~-\xi_m - \xi_{m'},~\xi_m - \xi_{m'}, -\xi_m + \xi_{m'}\},
\end{align}
with $\xi_m\geq 0$. From Eq.~\eqref{Wrates}, we find
\begin{equation}
        W^{\alpha\beta}_{mm}(\omega) = T_1\frac{1-i\omega T_1}{1+(\omega T_1)^2}\sum_k \delta A_k^{\alpha,z}\delta A_k^{\beta,z}\frac{V_m}{4}\frac{2Nm^2-1}{N-1},
\end{equation}
as well as
\begin{equation}
    W^{\alpha\beta}_{mm-1}(\omega) = V_{m} \frac{\nicefrac{N}{2} + m}{N} \left[\pi\varrho_{\alpha\beta}(\omega)-i~\text{p.v.}\int \dd\nu\frac{\varrho_{\alpha\beta}(\nu)}{\omega-\nu}\right],
\end{equation}
and
{
\begin{equation}
    W^{\alpha\beta}_{mm+1}(\omega) = V_{m} \frac{\nicefrac{N}{2} - m}{N} \left[\pi\varrho_{\alpha\beta}(-\omega)-i~\text{p.v.}\int d\nu\frac{\varrho_{\alpha\beta}(\nu)}{\omega+\nu}\right],
\end{equation}
}
where p.v. indicates the Cauchy principal value.
The real parts of these quantities determine the rates in the MARE while the imaginary parts contribute to the Lamb shift, see Eq.~\eqref{rates}. We may further simplify the MARE by considering spectral densities that vanish at low (and negative) frequencies, such that
\begin{equation}
    \varrho_{\alpha\beta}(-\xi_m-\xi_{m'})=\varrho_{\alpha\beta}[\pm(\xi_{m}-\xi_{m'})]=0,
\end{equation}
{which is valid for short-range jumps, $|m-m'|/N \ll 1$. In our particular case, $|m-m'|=1$, and we are typically interested in $N\gg 1$.}

Furthermore, we assume that the splitting of the eigenenergies of $H_m$ is large compared to the inverse $T_1$ time of the bath, such that
\begin{equation}
    \frac{1}{1+(\xi_m+\xi_{m'})^2T_1^2}\simeq 0.
\end{equation}
With these simplifications, we obtain the MARE
\begin{align}\label{genMAREapp} 
\partial_t\rho_{m} &= -i\qty[\vec{B}_m\cdot\vec{S}+H^{\rm LS}_m,
\rho_{m}] +\gamma_m\left(\mathcal{D}\left[\dyad{\uparrow_m}\right]+\mathcal{D}\left[\dyad{\downarrow_m}\right]\right)\frac{\rho_m}{V_m}\\ 
&+\Gamma_{m}\qty[\dyad{\downarrow_m}{\uparrow_{m-1}}\frac{\rho_{m-1}}{V_{m-1}}
\dyad{\uparrow_{m-1}}{\downarrow_m} -\frac{1}{2}\qty{\dyad{\downarrow_m}{\downarrow_m}, \frac{\rho_{m}}{V_{m}}}\nonumber]\\
&+\Gamma_{m+1}\qty[
\dyad{\uparrow_m}{\downarrow_{m+1}}\frac{\rho_{m+1}}{V_{m+1}}
\dyad{\downarrow_{m+1}}{\uparrow_{m}} -\frac{1}{2}\qty{\dyad{\uparrow_m}{\uparrow_m} ,
\frac{\rho_{m}}{V_{m}}}],\nonumber
\end{align}
with the rates
\begin{align}
\label{eq:ratesgen}
 \Gamma_m &= 2\pi V_m \left(\frac{1}{2}+\frac{m}{N}\right)\sum_{\alpha\beta}\Re{\varrho_{\alpha\beta}(\xi_m + \xi_{m-1}) F^{\alpha \downarrow \uparrow}_{m,m - 1}\qty(F^{\beta \downarrow \uparrow}_{m,m - 1})^*},\\
 \gamma_m & = 
 T_1\sum_k\sum_{\alpha\beta}\delta A_k^{\alpha,z}\delta A_k^{\beta,z}V_m\frac{N^2-4m^2}{4N(N-1)},
\end{align}
and the Lamb shift can be computed from Eqs.~\eqref{eq:lamb} and \eqref{rates}. To arrive at Eq.~\eqref{genMARE} in the main text, we drop the Lamb-shift Hamiltonian, and we set $\gamma_m=0$ by neglecting the $k$-dependence in the system-bath coupling.

{Note that our approach is to develop the bath correlation functions in terms of the bath spectral densities. In contrast, in Ref.~\cite{Riera2021}, the authors coarse grain over the eigenvalues of $M_B$—here, the magnetisation $m$, and in Ref.~\cite{Riera2021} bath energies—leading to decay in the bath correlation functions. Although both approaches are technically correct, the spectral density allows to include empirical bath decoherence processes and can be extracted from experimental data.}

\subsection{Large spin-$s$ bath}\label{a:sspin} 
We consider a bath with $N\gg
1$ identical spins and each spin has spin $s$. Counting
the possible states whose azymuthal projection add up
to a given value $m$ becomes impossible. However, an approximate value can be given in the large--$N$ limit, due to the
central limit theorem. To this end, we use a probabilistic approach: we consider the infinite temperature state, where each micro-state is equally likely. We then consider the probability, $p_m$, that we find a given magnetisation, $m$. The volume factor then reads
\begin{align}
    V_m = (2s+1)^N p_m,
\end{align}
where $(2s+1)^N$ is total number of possible configurations. In the infinite temperature state, the average and variance of a single spin are given by,
\begin{align} 
\mu = \expval{S_z} =\frac{1}{2s
+1}\sum_{j=-s}^s j =0,\hspace{2cm} \varsigma^2_s =
\text{var}{(S_z)} =\frac{1}{2s +1}\sum_{j=-s}^s j^2
=\frac{s(s+1)}{3}. 
\end{align} 
For non-interacting spins, the total magnetisation satisfies a central-limit theorem and we have
\begin{align}\label{sgauss} 
p_m(s)\approx 
\frac{3}{\sqrt{2\pi N} \varsigma_s }e^{-\frac{m^2}{N
\varsigma_s^2}} \end{align} 
which fully determines the volumes. This problem is equivalent to computing the microcanonical partition function of a Boltzmann gas~\cite{Salinas}.

As a consequence, the
spectral properties of a large ensemble of $N$ spin-$s$ nuclei can
be emulated by a larger ensemble of $N'$ spin$-1/2$ nuclei, as
long as both produce the same distribution.
Since Eq.~\eqref{sgauss} is fully determined by its variance, the condition
is,
\begin{align}\label{Nprime}
N' =\frac{4}{3} N s(s+1).
\end{align}
Thus, using this relation we can use all the results for
spin-$\nicefrac{1}{2}$ replacing $N$ by $N'$. In particular,
\begin{align}\label{general_Vm}
    V_m \approx \frac{[\nicefrac{4}{3} N s (s+1)]!}{[\nicefrac{2}{3} N s (s+1)+m]! [\nicefrac{2}{3} N s (s+1))-m]!}.
\end{align}

We now compute the rate in Eq.~\eqref{rate_plus}.  The calculation is possible by rewriting it fully in terms of the above volume factors, and thus holds under the the large--$N$ limit. To achieve this, we first, note that since Eq.~\eqref{rate_plus} has permutation invariance with respect to $k$, we can sum over $k$ and divide by $N$ without changing the result,
\begin{align}
    \tr{\Pi_m I_k^\dagger I_k} = \frac{1}{N} \tr{\Pi_m J^+ J},
\end{align}
where $J = \sum_k I_k$ are total spin operators. We now proceed by performing the calculation in the Dicke basis, in which the projectors write,
\begin{align}
    \Pi_m = \sum_{j=|m|}^{s N}\sum_{d=1}^{d_j} \dyad{j m d},
\end{align}
where $\vec{J}\cdot \vec{J} \ket{j m d} = j(j+1)  \ket{j m d}$, and  $d$ are degeneracies of each Dicke state due to permutations with the same $j,m$. Using standard relations from angular momenta,
\begin{align}
    \tr{\Pi_m J^\dagger J} = \sum_{j'=|m|}^{s N} d_j[j(j+1) - m (m - 1)].
\end{align}
The degeneracies of the Dicke states satisfy the property, 
\begin{align}
\sum_{j=|m|}^{sN}d_j = V_m.
\end{align} 
For a detailed discussion on the degeneracy of Dicke states, see Ref.~\cite{MandelWolf95}.
This allows the second term in the square brackets above to be summed directly, and implies that $d_j = V_{j} - V_{j+1}$. We thus have,
\begin{align}
\label{eq:sumeq}
\tr{\Pi_m J^\dagger J} &= V_m \sum_{j=|m|}^{s N} (V_j - V_{j+1})\qty[j(j + 1) - m (m-1)]
\end{align}
So far, we have not performed any approximation, and the steps do not depend on the value of $s$. The calculation is finished by using the approximate expression for the the volumes, Eq.~\eqref{general_Vm}, and we have,
\begin{align}
V_{j} - V_{j+1} &= \frac{(2j +1)N'!}{(\nicefrac{N'}{2} + j +1)!(\nicefrac{N'}{2} - j)!},\hspace{1cm}j\leq \frac{N'}{2}.
\end{align}
With the above expression we can perform the sum in Eq.~\eqref{eq:sumeq}.
As expected, the result is the same as for $s=1/2$, but with $N\to N' = N(4s/3)(s+1)$,
\begin{align}
r^{+}_{mm'}(s) &= \delta_{m',m-1} V_{m}
\qty[\frac{2}{3}s(s+1) + \frac{m}{N}],
\end{align}
resulting in the rates
\begin{align}
\label{eq:ratesgenls}
 \Gamma_m &= 2\pi V_m \qty[\frac{2}{3}s(s+1) + \frac{m}{N}]\sum_{\alpha\beta}\Re{\varrho_{\alpha\beta}(\xi_m + \xi_{m-1}) F^{\alpha \downarrow \uparrow}_{m,m - 1}\qty(F^{\beta \downarrow \uparrow}_{m,m - 1})^*}.
\end{align}
The same rates have been found in Ref.~\cite{Yan2016} via a heuristic argument.

\subsection{Conserved quantity}\label{a:conserved-quantity}
We now show that Eq.~\eqref{genMARE} conserves the quantity,
\begin{align}
M = m + \hat{B}_m \cdot \vec{S}.
\end{align}
Note that even the average of this quantity, $\E{M}$, in general contains correlations between system and bath, since $\hat{B}_m\cdot\vec{S}$ is an $m$-dependent operator. To show the conservation law, we observe that
\begin{align}
    \hat{B}_m \cdot \vec{S} \ket{\uparrow_m\!/\downarrow_m} = \pm \frac{1}{2}\ket{\uparrow_m\!/\downarrow_m}, ~\forall m,
\end{align}
i.e., the eigenvalues do not depend on $m$. With this observation, we can follow a similar argument to Ref.~\cite{Riera2021}: we show that the equation for the populations, $\partial_t p(\uparrow/\downarrow, m)$, implies that $\partial_t P(M) = 0$, with  
 \begin{align}
     P(M) = \sum_{m} \sum_{\sigma = \pm1/2}\delta_{M, m + \sigma} p(\sigma, m),
\end{align}
where $p(1/2, m) = p(\uparrow,m)$ and $p(-1/2, m) = p(\downarrow,m)$. Using Eqs.~\eqref{pdwm} and \eqref{pupm}, we find
\begin{align}
    \partial_t P(M) =&
    \Gamma_{M-1/2}\frac{p(\uparrow, M-1/2)}{V_{M-1/2}} -\Gamma_{M+1/2}\frac{p(\downarrow, M+1/2)}{V_{M-1/2}}
   \\& + \Gamma_{M + 1/2}\frac{p(\downarrow, M+1/2)}{V_{M-1/2}} -\Gamma_{M - 1/2}\frac{p(\uparrow, M-1/2)}{V_{M-1/2}}=0.\nonumber
\end{align}
    
\subsection{Analytic solution}\label{ss:analytic-sol}
We follow Ref.~\cite{Riera2021} and use the conserved quantity to construct the propagator acting on the populations, Eqs.~(\ref{pdwm},~\ref{pupm}). The evolution of the coherent part is simple and does not couple to the populations---we leave it for last. We emphasise that the equations of populations and coherences in general only decouple in the $m-$dependent basis spanned by $\{\ket{\sigma_m}\}_{\sigma, m}$.
The conservation of $M$ implies that the jump processes allowed by the MARE must conserve $M$, implying that the propagator can be decomposed as 
\begin{align}
    K = \bigoplus_{M} K_{|M},
\end{align}
where $K_{|M}$ is the restriction of $K$ to the subspace with a fixed $M$; these are two-dimensional subspaces spanned by pairs of the form, ${p(\uparrow, m),~ p(\downarrow, m+1)}$. To make progress, we introduce the vector,
\begin{align}
    \vec{p} = \begin{pmatrix}
    \vdots\\
    p(\uparrow,m-1)\\
    p(\downarrow,m)\\
    p(\uparrow,m)\\
    p(\downarrow,m+1)\\
    \vdots
    \end{pmatrix},
\end{align}
and write Eqs.~(\ref{pdwm}, \ref{pupm}) as $\partial_t \vec{p} = G \vec{p}$, where
where $G = \bigoplus_M G_{|M}$ with
\begin{align}
    G_{|M} = \frac{\Gamma_{m+1}}{V_m} 
    \begin{pmatrix}
      -1 & \frac{V_m}{V_{m+1}}\\
      1 & -\frac{V_m}{V_{m+1}}
    \end{pmatrix}.
\end{align}
Then, we simply have to exponentiate each block
\begin{align}
    K_{|M} =e^{t G_{|M}}=\mathds{1}_{|M} + \frac{1-e^{-t\bar{\Gamma}_m}}{\bar{\Gamma}_m}G_{|M}
    = \mathds{1}_{|M} + \frac{1-e^{-t\bar{\Gamma}_m}}{1 + V_{m}/V_{m+1}}
    \begin{pmatrix}
        -1 & \frac{V_m}{V_{m+1}}\\
        1  & -\frac{V_m}{V_{m+1}}
    \end{pmatrix}\nonumber,
\end{align}
where we introduced,
\begin{align}
    \bar{\Gamma}_m = \frac{\Gamma_{m+1}}{V_m}\qty( 1 +  \frac{V_m}{V_{m+1}}).
\end{align}
In the long-time limit, this propagator implies that the populations in the $m$-dependent basis must satisfy
\begin{align}\label{steady-state-condition}
    \frac{p_\text{ss}(\downarrow, m+1)}{p_\text{ss}(\uparrow, m)} = \frac{V_{m+1}}{V_m}.
\end{align}

We now discuss the coherent sector; $q(\sigma, \nu, m) = \bra{\sigma_m}  \rho_m\ket{\nu_m}$. From the MARE~\eqref{genMAREapp},
\begin{align}
    \partial_t q(\sigma, \nu, m) = -\qty[2i\xi_m + \frac{\gamma_m+\Gamma_m + \Gamma_{m+1}}{2V_m} ]q(\sigma, \nu, m).
\end{align}
The above equation can be directly exponentiated,  and we are done constructing the analytic solution. 

The plots in the main text are obtained by implementing the solutions numerically and evolving initial conditions. For the spin qubit, we use a cutoff for $|m|$ equal to $N/50$ to accelerate the calculations. This is justified since $P_m$ is never polarised excessively and leads to errors of the order of $10^{-10}$. {For the superconducting qubit, no cutoff is used}.

\subsection{Second law of thermodynamics}\label{a:second-law}
The existence of a conserved quantity also allows us to generalise the second law of thermodynamics from Ref.~\cite{Riera2021} to the MARE. We consider the observational entropy
\begin{align}\label{obs-entropy-pops}
    S_\text{obs} = \sum_{\sigma, m} p(\sigma, m)[-\ln p(\sigma, m)  + \ln V_m].
\end{align}
In this appendix, we prove that
\begin{align}
    \partial_t S_\text{obs} \geq 0.\label{second-law}
\end{align}
The crucial step is to show that the above is minus the  time-derivative of the Kullback-Leibler divergence between the generic probability vector $\vec{p}$ and  the corresponding steady-state vector, $\vec{p}_\text{ss}$;
\begin{align}
    \partial_t D_\text{KL}(\vec{p}~|~\vec{p}_\text{ss})= \partial_t\sum_{\sigma m} p(\sigma, m) \ln\frac{p(\sigma, m)}{p_\text{ss}(\sigma, m)}.
\end{align}
It sufficies to concentrate on,
\begin{align}
    &\partial_t \sum_{\sigma,m} p(\sigma, m)\ln p_\text{ss}(\sigma, m)
   =\partial_t\sum_{m}{\qty[p(\uparrow, m) \ln p_\text{ss}(\uparrow, m) + p(\downarrow, m) \ln p_\text{ss}(\downarrow, m)]}\\
    &=\partial_t\sum_{m}{\qty[P(M = m+\nicefrac{1}{2}) - p(\downarrow, m+1)\ln p_\text{ss}(\uparrow, m)\nonumber+ p(\downarrow, m) \ln p_\text{ss}(\downarrow, m)]}\\
    &=\partial_t\sum_{m} {\qty[p(\downarrow, m) \ln p_\text{ss}(\downarrow, m) - p(\downarrow, m+1)\ln p_\text{ss}(\uparrow, m)]} \nonumber,
\end{align}
where we used that $P(M) = \sum_{\sigma,m}\delta_{M, m + \sigma} p(\sigma, m),~\partial_t P(M) = 0,~\partial_t p_\text{ss}(\sigma, m) = 0$. We now use Eq.~\eqref{steady-state-condition}
\begin{align}
    &\partial_t \sum_{\sigma,m} p(\sigma, m)\ln p_\text{ss}(\sigma, m)= \partial_t\sum_{m}p(\downarrow, m) \ln p_\text{ss}(\downarrow, m)\\
    &-\partial_t \sum_m p(\downarrow, m+1) \ln p_\text{ss}(\downarrow, m+1) -\partial_t \sum_m p(\downarrow, m+1) \ln \frac{V_{m+1}}{V_m}.\nonumber
\end{align}
The first two terms cancel each other upon change of indices and we are left with,
\begin{align}
&-\partial_t \sum_{\sigma,m} p(\sigma, m)\ln p_\text{ss}(\sigma, m)=-\partial_t \sum_m\qty[p(\downarrow, m+1) \ln V_{m+1} - p(\downarrow, m+1) \ln V_{m}]\\
&=-\partial_t \sum_m\qty[p(\downarrow, m) \ln V_{m} + p(\uparrow, m) \ln V_{m}]=-\partial_t \sum_{\sigma, m} p(\sigma,m ) \ln V_m,\nonumber
\end{align}
where we used again the definition of $P(M)$ and the conservation law, $\partial_t P(M)=0$. 
With the result above, we have that
\begin{align}
    \partial_t D_\text{KL}(\vec{p}|\vec{p}_\text{ss}) =  - \partial_t S_\text{obs}
\end{align}
To complete the proof, we note that the dynamics of $\vec{p}$ is Markovian, see Eqs.~(\ref{pdwm}, \ref{pupm}). Then, we can use the data-processing inequality~\cite{CoverThomas2005}, which implies
\begin{align}
    -\partial_t D_\text{KL}(\vec{p}|\vec{p}_\text{ss})  \geq 0,
\end{align}
and we have shown that Eq.~\eqref{second-law} is satisfied.

\subsection{Decomposing the observational entropy}\label{a:decomposition}
Here we consider decompositions of the observational entropy. 
To this end, we note that it can be split in terms of marginal entropies and coarse-grained mutual information~\cite{Riera2021},
\begin{align}
    \mathcal{S}_\text{obs} &= \mathcal{S}_B(P) + \mathcal{S}(p_S) 
            - D_{\text{KL}}[\vec{p} |P~p_S]
        + \sum_m P_m \ln V_m.
\end{align}
We further note that $\mathcal{S}(p_S)$---the Shannon entropy of the marginal $p_S(\sigma) = \sum_m p(\sigma, m)$---does not bear an obvious relation with the von Neumann entropy of the system.  It can however be associated to the von Neumann entropy of a generalised dephasing map, $\mathscr{D}$, acting on $\rho_S$,
\begin{align}
    \mathscr{D}\rho_S =  \sum_m \mathscr{D}_m(\rho_m)
    = \sum_m \frac{1}{2\pi} \int_0^{2\pi} \dd \chi e^{i \chi \hat{B}_m \cdot \vec{S}} \rho_m e^{-i \chi \hat{B}_m \cdot \vec{S}}.
\end{align}
We then write,
\begin{align}
    \mathcal{S}(p_S) = S(\rho_S) + \mathcal{C}_\text{rel},
\end{align}
where we introduced the (generalised) relative entropy of coherence~\cite{Streltsov2017}, 
\begin{align}
\mathcal{C}_\text{rel} = \mathcal{S}(\mathscr{D}\rho_S) - \mathcal{S}(\rho_S).
\end{align}
We can then write the observational entropy as,
\begin{align}
    \mathcal{S}_\text{obs} &=\mathcal{S}_B(P) + \mathcal{S}(\rho_S)   + \mathcal{C}_\text{rel}
             - D_{\text{KL}}[\vec{p} |P~p_S] +\sum_m P_m \ln V_m.
\end{align}
Finally, we argue that $\mathcal{C}_\text{rel}\geq 0$.
Note that $\mathscr{D}$ is CPTP and unital, $\mathscr{D}(\mathds{1})=1$. Using the monotonicity of the quantum relative entropy under CPTP maps~\cite{Lindblad1975},
\begin{align}
     D_\text{KL}(\rho_S|\mathds{1}) \geq D_\text{KL}\qty[\mathscr{D}(\rho_S)|\mathscr{D}(\mathds{1})],\hspace{1cm}
    \Rightarrow \hspace{1cm}  \mathcal{S}(\mathscr{D}\rho_S) \geq \mathcal{S}(\rho_S),
\end{align}
Therefore, $\mathcal{C}_\text{rel}\geq 0$. That is, the von Neumann entropy of the system discsussed in the main text lower bounds the entropy of the marginal,
\begin{align}
    \mathcal{S}({p_S})\geq \mathcal{S}(\rho_S).
\end{align}

\section{Superconducting qubit}\label{a:sc}
Below we derive Eq.~\eqref{sc-MARE} by providing an effective Hamiltonian  and specialising the results of App.~\ref{a:central-spin-mare}.

\subsection{Schrieffer-Wolff transformation}\label{a:SW}

In this subsection we consider TLSs with arbitrary frequencies. Via Schrieffer-Wolff pertubation theory we show that the far off-resonant TLSs do not contribute to the magnetisation exchange processes and justify concentrating on the quasi-resonant TLSs. 
We consider the Hamiltonian $H=H_0+V$ with
\begin{equation}
     H_0 =
\omega_SS_z+\sum_{k=1}^{N_{\rm tot}}\omega_kS_z^k,
\end{equation}
and the coupling Hamiltonian
\begin{align}\label{sc-interaction1} V =  \sum_{k=1}^{N_{\rm tot}} A_k
\Vec{S}\cdot\Vec{S}_k=\sum_{k=1}^{N_{\rm tot}} A_k S_z S_z^k + \frac{1}{2}\sum_{k=1}^{N_{\rm tot}}
A_k \qty(S^\dagger S_k + S_k^\dagger S),
\end{align} 
with $\Vec{S}_{(k)}=(S^{(k)}_x, S^{(k)}_y, S^{(k)}_z)$. We consider cases in which
$V$ represents \textit{weak} coupling, with hyperfine constants $A_k$.
The  TLSs ensemble can be decomposed into two
parts: the quasi-resonant TLSs that have a frequency close to $\omega_S$ and the TLSs that are off-resonant. The set of quasi-resonant TLSs is defined by
\begin{align}
O_{\delta} =
\{\omega_k:~|\omega_k-\omega_S|<\delta\}.\label{quasi-resonant}
\end{align}
The off-resonant TLSs have frequencies $\omega_k\notin O_\delta$ that are farther than
$\delta$ away from $\omega_S$. Following Ref.~\cite{landi2024-SW}, we may then apply a Schrieffer-Wolff transformation to reduce the interaction to 
\begin{align}\label{Vsc} V \approx  \sum_{k:\omega_k\in O_\delta} A_k
\Vec{S}\cdot\Vec{S}_k + \sum_{k:\omega_k\notin
O_\delta}\frac{A_k^2}{\Delta_k}S_z^k S_z, \end{align} where $\Delta_k =
\omega_k-\omega_S$. Assuming the off-resonant TLS to be in a state that is diagonal in the basis of $S_z^k$, their effect on the system is fully captured by a random but time-independent shift in the frequency $\omega_S$. Since we do not investigate the coherence properties of the superconducting qubit, this has no effect on our results and we drop it in the following.

\subsection{MARE for the superconducting qubit}\label{a:SC-deriv}

We start by writing the Hamiltonian obtained from the Schrieffer-Wolff pertubation theory as,
\begin{align}
    H &= H_S + \delta H + H_B + V,\\
    H_S = \omega_S S_z,\hspace{.5cm}
    \delta H  &= a J_z,\hspace{.5cm}
    V = \sum_{k=1}^N \delta A_k S_z S_k^z + \sum_{k=1}^NA_k\qty(S_x S_k^x + S_y S_k^y),
\end{align}
where we introduced $a = \sum_{k=1}^N A_k/N,~\delta A_k = A_k - {a}$. Here, all $A_k$ are perturbative and the sum runs only over quasi-resonant TLSs. The term associated to $\delta A_k$  is small even compared with $A_k$, since we consider only quasi-resonant TLSs. This justifies discarding its contributions to the MARE. Thereby, the $m$-dependent Hamiltonian is simply,
\begin{align}
    H_m = \vec{B}_m \cdot \vec{S} = (\omega_S + am) S_z.
\end{align}

This Hamiltonian always leads to jumps between $S_z$ eigenstates, and the jump operators will not depend on $m$ and the MARE has the form,

\begin{align}
    \partial_t\rho_{m} = -i\qty[\vec{B}_m\cdot\vec{S},
\rho_{m}] 
+\Gamma_{m}\qty[\sigma\frac{\rho_{m-1}}{V_{m-1}}
\sigma^\dagger -\frac{1}{2}\qty{\sigma\sigma^\dagger, \frac{\rho_{m}}{V_{m}}}]+\Gamma_{m+1}\qty[
\sigma^\dagger\frac{\rho_{m+1}}{V_{m+1}}
\sigma -\frac{1}{2}\qty{\sigma^\dagger \sigma ,
\frac{\rho_{m}}{V_{m}}}],\nonumber
\end{align}
where $\vec{B}_m = (\omega_S + Am)\hat{e}_z$. Above, the rates are given by,
\begin{align}
    \Gamma_m = V_m\left(\frac{1}{2}+\frac{m}{N}\right) \kappa_m,
\end{align}
with $\kappa_m=2\pi\varrho(\omega_S + am)$. Reminding that $A_k$ are perturbative and $a = \sum_k A_k/N$, we can approximate for the quasi-resonant TLSs $\varrho(\omega_S + am)\approx \varrho(\omega_S)$. The couplings $\kappa_m$ then become independent of $m$ and we recover Eq.~\eqref{sc-MARE}. In general, for the quasi-resonant TLSs, the non-perturbative contribution is small and can be discarded, as long as $|N a|\ll \omega_S$. If the off-resonant TLSs are also included, or the region of quasi-resonant TLSs is very dense, this approximation is inadequate.

{

\subsection{Details on variance reduction}\label{a:variance-reduction}

Equation~\eqref{sc-var-ss} serves as the basis to understand variance reduction in the cooling (uncorrelated) scenario and in the narrowing (correlated) protocol studied in Section~\ref{s:sc}. Furthermore, we also use it to derive the bounds on variance reduction mentioned therein.

\subsubsection{Thermal initial states}

For thermal states,
\begin{align}
\langle m\rangle_\beta &= -\frac{N}{2}\tanh\!\left(\frac{\beta\omega_S}{2}\right) =: -\frac{N}{2}t_\beta,\\
\langle\!\langle m^2\rangle\!\rangle_\beta &= \frac{N}{4}\sech^2\!\left(\frac{\beta\omega_S}{2}\right) = \frac{N}{4}(1-t_\beta^2),
\end{align}
where we used the quasi-resonant conditions $\omega_k \approx \omega_S$.
Substituting into Eq.~\eqref{sc-var-ss}, we obtain
\begin{align}
\langle\!\langle m^2 \rangle\!\rangle_\infty &=  \frac{N}{4}(1-t_\beta^2)\qty[\frac{N-1}{N+1} - \frac{1}{N(N+1)^2}\frac{(N t_\beta + 1)^2}{1 - t_\beta^2} + \frac{1}{N(1- t_\beta^2)}]\\
&=  \langle\!\langle m^2 \rangle\!\rangle_\beta \qty[\frac{N-1}{N+1} + \frac{(N+1)^2 - (N t_\beta + 1)^2}{N(N+1)^2(1- t_\beta^2)}]\nonumber\\
&=\langle\!\langle m^2\rangle\!\rangle_\beta 
\left[
\frac{N}{N+1}
+
\frac{1}{(N+1)^2}
\frac{1- t_\beta}{1 + t_\beta}
\right]\nonumber\\
&=\langle\!\langle m^2\rangle\!\rangle_\beta 
\left[
\frac{N}{N+1}
+
\frac{e^{-\beta \omega_S}}{(N+1)^2}
\right],
\nonumber\label{eq:thermal-variance-reduction}
\end{align}
where we used the appropriate conserved quantity provided by thermal states, an initial state in the qubit $\ket{\downarrow}$ and discarded correlations.

For large $N$, this simplifies to
\begin{align}
\langle\!\langle m^2\rangle\!\rangle_\infty
=
\langle\!\langle m^2\rangle\!\rangle_\beta
\left(1 - \frac{1}{N}\right)
+ \mathcal{O}\!\left(\nicefrac{1}{N}\right).
\end{align}
At the right-hand side of the above expression, note that the lowest order in the first term is $\mathcal{O}(1)$, since $\langle\!\langle m^2\rangle\!\rangle_\beta$ is $\mathcal{O}(N)$. 

\subsubsection{Correlated initial states.}
For both  $\Theta$-- and Ramsey--correlated initial states considered in Section~\ref{ss:sc_correlated},
\begin{align}
    \langle S_z \rangle_0 = 0,~\langle\!\langle S_z^2 \rangle\!\rangle_0 = 1/4,~\langle\!\langle S_z m \rangle\!\rangle = -k \varsigma,
\end{align}
where we remind that $\varsigma^2 = \langle \!\langle m^2 \rangle\!\rangle_0$ is the variance of the initial distribution $P_m = \mathcal{N}(0, \varsigma)$. With this boundary conditions, Eq.~\eqref{sc-var-ss} reduces to,
\begin{align}
    \langle\!\langle m^2 \rangle\!\rangle_\infty &= \frac{N-1}{N+1} \qty( \varsigma^2 - 2 k \varsigma + \frac{1}{4}) + \frac{1}{4}\\
    &= \varsigma^2 \qty( 1- \frac{2}{N}) - 2k \varsigma + \frac{1}{2} + \mathcal{O}(\nicefrac{1}{\sqrt{N}}),
\end{align}
where we emphasise that the large--$N$ expansion has been performed assuming $\varsigma \propto \sqrt{N}$. Equations~(\ref{eq:sc_var_reduction_Theta},~\ref{eq:sc_var_reduction_Ramsey}) have been obtained from the above expansion by setting $k=\sqrt{2\pi}^{-1}$ and $k = \sqrt{4e}^{-1}$ for $\Theta$-- and Ramsey--correlated cases, respectively.

\subsection{Conditions on variance reduction}\label{a:bounds-variance}

Equation~\eqref{sc-var-ss} can also be used to find strict conditions for variance reduction. Subtracting the initial variance from both sides of it and using the initial conditions to determine the conserved quantities, we obtain the change in variance,
\begin{align}
\delta \langle\!\langle m^2 \rangle\!\rangle
&=
\frac{1}{4}
- \frac{2}{N+1}\langle\!\langle m^2\rangle\!\rangle_0
+ \frac{N-1}{N+1}\qty(\langle\!\langle S_z^2\rangle\!\rangle_0 + 2\langle\!\langle S_z m\rangle\!\rangle_0)
- \frac{1}{(N+1)^2}(\langle m\rangle_0 - \langle S_z\rangle_0)^2. \label{eq:delta-var}
\end{align}
Below we analyse the allowed initial variance such that $\delta \langle\!\langle m^2 \rangle\!\rangle =  \langle\!\langle m^2 \rangle\!\rangle_\infty - \langle\!\langle m^2 \rangle\!\rangle_0 <0 $, that is, we find the thresholds $\varsigma^*$,
\begin{align}
    \varsigma>\varsigma^* ~\Leftrightarrow~ \delta\langle\!\langle m^2 \rangle\!\rangle < 0,
\end{align}
for each case of interest.

\subsubsection{Uncorrelated states}
Here, we concentrate on the protocol with $\langle S_z \rangle_0 = -1/2, \langle\!\langle S_z \rangle\! \rangle_0 = 0$, and $\langle\!\langle S_z m\rangle\!\rangle_0=0$. We obtain from $\delta \langle\!\langle m^2 \rangle\!\rangle  < 0$,
\begin{align}
\langle\!\langle m^2\rangle\!\rangle_0
>
\frac{N+1}{8}
-
\frac{1}{2(N+1)}(\langle m\rangle_0 - 1/2)^2.
\end{align}
In particular, for an unpolarised bath, $\langle m\rangle_0 = 0$, 
\begin{align}
 \langle\!\langle m^2\rangle\!\rangle_0
>\frac{N}{8}\frac{N-2}{N+1} = \frac{N}{8} + \mathcal{O}(\nicefrac{1}{N}) = (\varsigma^*)^2.
\end{align}
The exact value in Table~\ref{t:sc-comparison} for $\varsigma_*$ correspoding $\ket{\downarrow}$  is computed from the above equation.

\subsubsection{Thermal condition}

Reminding that $t_\beta=\tanh(\beta\omega_S/2)$, the condition  $\delta\langle\!\langle m^2 \rangle\!\rangle <0$ becomes
\begin{align}
\frac{N}{4}(1-t_\beta^2)
>
\frac{N+1}{8}
-
\frac{(1-Nt_\beta)^2}{8(N+1)}.
\end{align}
This simplifies to
\begin{align}
(N+2)(1+t_\beta)>2,
\end{align}
i.e.
\begin{align}
t_\beta>-\frac{N}{N+2}.
\end{align}
Thus,
\begin{align}
\beta>-\frac{1}{\omega_S}\ln(N+1),
\end{align}
where we used that $\tanh^{-1}(x)$ is monotonic increasing and $\tanh^{-1}(x) = \ln[(1+x)/(1-x)]/2$.
Therefore, for finite--$\beta$  variance always reduces. A reciprocal result can be obtained for the protocol with $\ket{\uparrow}$ as an initial state, which would lead to the same variance reduction, but for negative-temperature thermal states.

\subsubsection{Correlated states}
For the correlated states, we get from Eq.~\eqref{eq:delta-var},
\begin{align}
\varsigma^2 + k(N-1)\varsigma - \frac{N}{4} > 0
\end{align}
where we considered, 
\begin{align}
\langle m\rangle_0=0,\qquad \langle S_z \rangle_0 = 0, \qquad \langle\!\langle S_z m\rangle\!\rangle_0=-k\varsigma, \qquad \varsigma^2=\langle\!\langle m^2\rangle\!\rangle_0,
\end{align}
which hold for $\rho_m^{\Theta/R}$.

For $N>2$, we find the solution
\begin{align}
\varsigma >
-\frac{
 k (N - 1)}{2}
+ \frac{1}{2}\sqrt{
k^2 (N - 1)^2
+ N
} = \varsigma^*.
\end{align}
And in the large-$N$ limit,
\begin{align}
\varsigma > \frac{1}{4k} + \mathcal{O}\!\left(\nicefrac{1}{N}\right),
\end{align}
leading to
\begin{align}
&\varsigma > \sqrt{\frac{\pi}{8}}, \qquad (\Theta~\text{state}),\\
&\varsigma > \sqrt{\frac{e}{4}}~(\text{Ramsey state}).
\end{align}
The values related to $\rho_m^\Theta, ~\rho_m^R$ in the last column of Table~\ref{t:sc-comparison} are obtained from the above equations.
Note that since $\langle S_z \rangle_0 = 0$ for the correlated states, setting $k = 0$ above does \textit{not} correspond to the uncorrelated protocol above.
}

\section{Spin qubit}\label{a:sq}

\subsection{MARE for the spin qubit}
For the spin qubit, we start from the Hamiltonian in Eq.~\eqref{hamqd}. From Eq.~\eqref{eq:hamm}, we obtain the $m$-resolved Hamiltonian,
\begin{align}
    H_m = \vec{B}_m \cdot \vec{S},\hspace{2cm}
    \vec{B}_m = (\Delta+A_{\rm c}m)\hat{e}_z+\Omega\hat{e}_x.
\end{align}
This Hamiltonian has the eigenstates
\begin{align}
    \ket{\lambda_m} &=\sin (\chi^{\lambda}_m) \ket{\downarrow_z} + \cos (\chi^{\lambda}_m) \ket{\uparrow_z},~\lambda=\downarrow,\uparrow\\
    \chi^{\uparrow (\downarrow)}_m &= \tan^{-1}\frac{{2}\xi_m\pm\Delta_m }{\Omega},\hspace{2cm}
    \xi_m =\frac{|\vec{B}_m|}{2}= \frac{1}{2}\sqrt{\Delta_m^2 + \Omega^2},
\end{align}
where we used the shorthand, $\Delta_m = \Delta  +A_{\rm c}m$. To find the rates, we also need
\begin{align}
   |F^{\downarrow \uparrow}_{mm'}|^2=  \qty|\bra{\downarrow_m}S_z\ket{\uparrow_{m'}}|^2=\frac{\qty[(2\xi_m-\Delta_m)(2\xi_{m'}+\Delta_{m'}) + \Omega^2]^2}{64 \xi_m\xi_{m'}(2\xi_m-\Delta_m)(2\xi_{m'}+\Delta_{m'})}.
\end{align}
The rates in the final MARE are then given by, see Eq.~\eqref{eq:ratesgenls},
\begin{align}
    \Gamma_m =  2\pi V_m\left[\frac{2}{3}s(s+1) +  \frac{m}{N}\right] |F^{\downarrow \uparrow}_{m,m-1}|^2\varrho\qty(\xi_m +\xi_{m-1}).\label{eq:Gamma_m_expression}
\end{align}

We assume that the spectral density is Lorentzian, motivated by an exponential decay of the correlation functions at high Zeeman field~\cite{Coish2008decay, Wust2016, Jackson2022},
\begin{align}
    \varrho(\omega)  = 4A_\text{nc}^2 N \frac{\gamma}{(\omega_B - \omega)^2 + \gamma^2}.
\end{align}

We observe that this form of the spectral density is a simplification. In spin qubits, the dominant nuclear dephasing leading to the decay of the bath correlation functions  can be an electron(qubit)-mediated effect which demands the inclusion of higher order contributions of inhomegeneous flip-flop processes~\cite{Coish2004, Coish2008decay, Wust2016, Barnes2011} and strain-mediated quadrupole processes~\cite{Gangloff2019, Zaporski2023}. Furthermore, note that multiple nuclear substances and isotopes can be included by modifying the rates and including multiple peaks in the spectral densities at their characteristic frequencies. Below, we discuss the regime of validity of our approximations for typical experimental parameters.

\subsection{Validity of the approximations}\label{a:approximations}
The Markov approximation is justified whenever the
 bath correlation time, $\tau_B=1/\gamma$, { is small compared to the relaxation time of $\rho_m$, $\tau_m$ . In practice, the relaxation times $\tau_m$ can be estimated by,
\begin{align}
    \tau_m = \frac{V_m}{\Gamma_m}\geq \tau_R= \frac{V_0}{\Gamma_{0}}.
\end{align}
To obtain a conservative estimate for $\tau_B<\tau_R$, we assume Hartmann-Hahn resonance,
$\Omega= \omega_B$  where $\Gamma_m$ attain their largest values and we can estimate a tight upper bound. For this, in Eq.~\eqref{eq:Gamma_m_expression},
we use $\xi_{0} \approx \Omega/2, \xi_{\pm1} \approx \Omega/2 \Rightarrow |F_{0,-1}^{\downarrow \uparrow}|^2 \varrho(\Omega) \approx  A_\text{nc}^2/\gamma$. 
Thus, 
\begin{align}
     \tau_\text{R}^{-1} \lesssim 
\frac{4\pi}{3} s(s+1) N \frac{A_\text{nc}^2}{\gamma}.
\end{align}
Therefore, a safe condition for the Markov approximation is ($\tau_B/\tau_R < 1$), is
\begin{align}
\frac{4\pi}{3} \frac{A_\text{nc}^2N}{\gamma^2}s(s+1) \ll 1.
\end{align}
}

The secular approximation discards terms in which the jump terms contain frequencies $\omega \neq  \omega'$. The condition for the secular approximation is that  oscillations with (difference) frequencies $|\omega-\omega'|$ are fast compared to the system relaxation time; i.e., $|\omega - \omega'|^{-1}\ll
\tau_\text{R}$. As long as the spectral density is suppressed at frequencies close to zero (i.e., for the frequencies $|\xi_m-\xi_{m'}|$, only the frequencies $\omega = \xi_m+\xi_{m'}$ and $\omega' = -\xi_m-\xi_{m'}$ are relevant [see Eq.~\eqref{eigenfreqs}].
We have at Hartmann-Hahn resonance,  $|\omega - \omega'|\geq 4\min\{\xi_{m}, \xi_{m'}\}\geq { 2}\Omega$. A conservative estimate for the validity of the secular approximation is then,
\begin{align}
    { \frac{2\pi}{3}\frac{A_\text{nc}^2 N}{\gamma \Omega}s(s+1)\ll 1.}
\end{align}

\end{document}